\documentclass[12pt]{iopart}

\usepackage{iopams}  
\usepackage{units}
\usepackage{textcomp}
\usepackage{amstext}
\usepackage{amssymb}
\usepackage{stmaryrd}
\usepackage{graphicx}

\begin{document}

\title{Interaction of two Rydberg atoms in the vicinity of an optical nanofibre}

\author{E. Stourm$^{1}$, M. Lepers$^{2}$, J. Robert$^{1}$, S. Nic Chormaic$^{3}$, K. Mølmer$^{4}$, E. Brion$^{5}$}

\address{$^{1}$Université Paris-Saclay, CNRS, Laboratoire de physique des gaz et des plasmas, 91405, Orsay, France

$^{2}$Laboratoire Interdisciplinaire Carnot de Bourgogne, CNRS, Université de Bourgogne Franche-Comté, 21078 Dijon, France

$^{3}$Okinawa Institute of Science and Technology Graduate University, Onna, Okinawa, 904-0495, Japan

$^{4}$Niels Bohr Institute, University of Copenhagen, Blegdamsvej 17, 2100 Copenhagen, Denmark

$^{5}$Laboratoire Collisions Agrégats Réactivité, IRSAMC and UMR5589 du CNRS, Université de Toulouse III Paul Sabatier, F-31062 Toulouse Cedex 09, France}
\ead{brion@irsamc.ups-tlse.fr}
\vspace{10pt}

\begin{abstract}
  We consider two rubidium atoms, prepared in the same S or P Rydberg states  near an optical nanofibre. We determine the van der Waals interaction between them and identify novel features, including the reshaping of the interaction anisotropy and the formation of an interaction potential well near the nanofibre for P states. We attribute these phenomena to the breaking of the rotation symmetry  around the interatomic axis due to the presence of the fibre. Our work constitutes an important step in the assessment of Rydberg atom-nanofibre quantum interfaces and devices. 
\end{abstract}

%
%
%
%
%

\section{Introduction}
Interfacing atomic ensembles with light in a quantum network is a promising
way to achieve scalability of quantum architectures and devices, one of the crucial challenges in quantum technologies. Photons are ideal messengers between the atomic nodes of such a network. Protocols considered so far include free-space setups \cite{HSP10,LST09,CRF05,SSR11}, which are relatively easy to implement but suffer the drawback of strong losses. Optical nanowaveguides, in particular optical nanofibres (ONFs), constitute an interesting alternative. which offer strong transverse confinement of the field \cite{CDG18} and hence strong coupling. ONFs received much attention within the past two decades \cite{NGN16,SGH17}. For instance, the coupling to evanescent guided modes was used to trap \cite{BHK04,GET22}
and detect atoms \cite{NMM07} near a nanofibre. It was also theoretically shown that energy can be exchanged between two distant atoms via these modes \cite{LDN05}.

Within the past two decades, the strong dipole-dipole interaction
experienced by two neighbouring Rydberg-excited atoms and the associated so-called Rydberg blockade phenomenon \cite{LFC01} became the
main ingredient for many atom-based quantum information protocol proposals \cite{SWM10}, including atomic quantum registers \cite{BMS07} and repeaters \cite{BCA12}. Recently, preliminary steps were taken towards building a quantum network based on Rydberg-blockaded atomic ensembles linked via an optical nanofibre. The excitation of cold $^{87}$Rb atoms towards Rydberg $29\mbox{D}$ state was thus experimentally demonstrated at submicron distances from an optical nanofibre surface in a two-photon ladder-type excitation scheme \cite{RRK20}. 

On the theory side, the spontaneous emission of a highly excited sodium atom in the neighbourhood of a silica optical nanofibre was investigated \cite{SZL19}. The dependence of the emission rates into the guided and radiative modes on the radius of the fibre, the distance of the atom to the fibre, and the symmetry of the Rydberg state was studied. 
Since it used the so-called mode function approach, this work did not account for the fibre's absorption and dispersion. This is critical for Rydberg atoms that can de-excite along transitions of different frequencies for which the fibre index is different and potentially complex. Hence, the framework of Macroscopic Quantum Electrodynamics (MQE)  \cite{Buh12} has been employed to study a Rydberg-excited $^{87}$Rb atom near a silica nanofibre \cite{SLR20}. In MQE, one can take the exact refractive index of silica into account, thereby relaxing all constraints on addressable transitions. MQE also offers a natural way to compute not only atomic spontaneous emission rates but also Lamb shifts, which are modified by the presence of the nanofibre when compared to free-space, as was recently shown for low-lying excited levels of alkali-metal atoms \cite{KKN22}. As $n$ increases, the contribution of quadrupolar transitions to Lamb shifts and associated dispersion forces becomes important, as previously established for Rydberg atoms near metallic surfaces \cite{CEC10}. This contrasts with spontaneous emission rates for which quadrupolar transitions have negligible influence. 
Moreover, as already noticed for low-excited atoms \cite{LR14}, spontaneous emission may become directional when an atom is prepared in an excited angular momentum eigenstate defined relative to a quantisation axis which differs from the fibre axis. This effect is due to the peculiar polarisation structure of the field in the neighbourhood of the fibre. It is particularly strong for photons emitted into the fibre-guided modes and persists even for high principal quantum numbers, $n$. This is promising in view of potential applications in chiral quantum information protocols \cite{LMS17} based on a Rydberg atom-nanofibre interface.

Giant van der Waals interactions are among Rydberg atoms' most striking
features. In free-space, for two atoms prepared in levels of principal quantum numbers
$n>50$ and a few $\textrm{\ensuremath{\mu}m}$ apart, such interactions
can indeed induce energy shifts of the order of tens of $\mathrm{GHz}$.
In this scenario, the potential between two atoms $\left(A,B\right)$
separated by the distance $r_{AB}$ follows the law identified by
London \cite{Gal94}
\[
U_{AB}^{\left(0\right)}=-\frac{C_{6}\left(A,B\right)}{r_{AB}^{6}}
\]
The $C_{6}$ coefficient depends on the states in which the atoms $\left(A,B\right)$ are prepared as well as their geometric arrangement. It scales with the  principal quantum number as $n^{11}$. For a pair of rubidium atoms in the state $|60S_{1/2}\rangle$ in free-space it is of the order of $100~\mathrm{GHz}.\left(\mu\mathrm{m}\right)^{-6}$. 

In this article,  we investigate how the presence of the fibre modifies this interaction
with respect to the free-space case. This study follows
other works in plane geometries involving Rydberg atoms in
front of a conducting half-space \cite{BS17}.
In Sec. \ref{Presentation}, we first present the system under consideration, fix notations and specify the hypotheses we make. In particular, we briefly recall the form of the interaction Hamiltonian between two Rydberg atoms in the presence of a dieletric medium. 
In Sec. \ref{IntSS}, we study how the presence of the fibre modifies the interaction potential between two atoms prepared in the same state $|nS_{1/2}\rangle$, with $n\geq30$, in specific geometric configurations. In particular, we investigate how this potential evolves with the interatomic distance and the principal quantum number $n$. The novel features observed are attributed to the appearance of new couplings, forbidden in homogeneous free space but allowed by the fibre-induced symmetry breaking. 
In the case of two atoms prepared in the state $|nP_{3/2},M_{j}=\frac{3}{2}\rangle$ these new couplings may even dominate those existing in free-space and strongly enhance the potential, as we show in Sec. \ref{IntPP}. Due to the existence of a F{\"o}rster quasi-resonance, the interaction may also be strongly modified in its nature as $n$ increases. While the interaction is purely repulsive in free-space, we show that, in the vicinity of the nanofibre  and in certain geometric configurations, an interaction potential well can form. In Sec. \ref{Anisotropy}, we finally investigate how the interaction potential depends on the relative direction of the atomic orbital momenta to the interatomic axis in the presence of the fibre and compare to the case of free-space before concluding in Sec. \ref{Conclusion}.

\section{Presentation of the system, hypotheses and basic equations\label{Presentation}}

\begin{figure*}
\begin{centering}
\includegraphics[width=14cm]{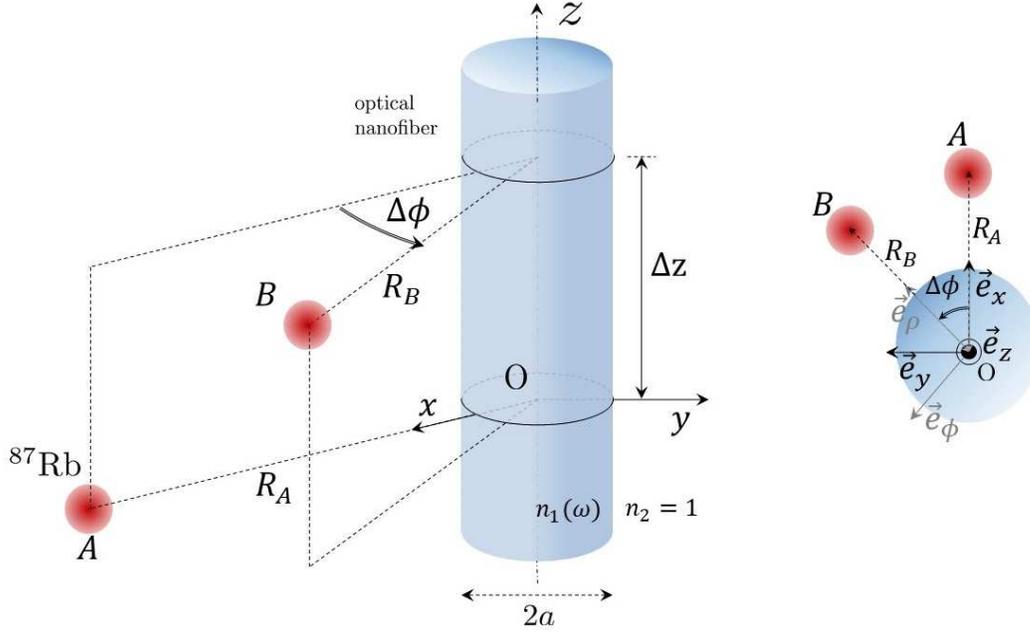}
\par\end{centering}
\caption{\textbf{Two  $^{87}\mbox{Rb}$ atoms, $\left(A,B\right)$,
near a silica optical nanofibre. Notations}. The Cartesian
frame $\left(Oxyz\right)$ is represented: i) its origin, $O$, is the projection of atom $A$'s centre of
mass on the fibre axis, ii) the $\left(Oz\right)$ axis coincides
with the fibre axis and is directed from atom $A$ towards $B$, iii)
the $\left(Ox\right)$ axis is along $\left(OA\right)$, and directed
from $O$ towards atom $A$, iv) the $\left(Oy\right)$ axis is chosen
so that $\left(Oxyz\right)$ is a direct frame. The unitary Cartesian
basis $\left({\bf e}_{x},{\bf e}_{y},{\bf e}_{z}\right)$ is represented
on the figure. A point $M$ of Cartesian coordinates $\left(x,y,z\right)$
is also identified by its cylindrical coordinates $\left(\rho\protect\geq0,0\protect\leq\phi<2\pi,z\right)$
defined by $\left(x=\rho\cos\phi,y=\rho\sin\phi,z\right)$. In particular,
atoms $A$ and $B$ have respective cylindrical coordinates $\left(R_{A},0,0\right)$
and $\left(R_{B},\Delta\phi,\Delta z\right)$. The local cylindrical
basis at point $M\left(\rho,\phi,z\right)$ is defined by the unit vectors ${\bf e}_{\rho}\equiv\cos\phi{\bf e}_{x}+\sin\phi{\bf e}_{y}$, ${\bf e}_{\rho}\equiv-\sin\phi{\bf e}_{x}+\cos\phi{\bf e}_{y}$.
The fibre radius is denoted by $a$.}
\label{SchemaDeuxAtomes}
\end{figure*}

We shall consider the idealised configuration represented in figure
\ref{SchemaDeuxAtomes}. Two rubidium atoms, $^{87}$Rb, denoted by
$A$ and $B$, respectively, are located near an infinite
cylindrical silica optical nanofibre of radius $a$. The Cartesian,
$\left(x,y,z\right)$, and cylindrical, $\left(\rho,\phi,z\right)$,
coordinates and associated bases, $\left({\bf e}_{x},{\bf e}_{y},{\bf e}_{z}\right)$
and $\left({\bf e}_{\rho},{\bf e}_{\phi},{\bf e}_{z}\right)$, are
defined in figure \ref{SchemaDeuxAtomes}. In particular, the
centres of mass of atoms $A$ and $B$ are identified by their cylindrical
coordinates $\left(R_{A},0,0\right)$ and $\left(R_{B},\Delta\phi,\Delta z\right)$,
respectively.

As we shall see, the van der Waals interaction between two Rydberg atoms are mainly due to transitions between the initial state and close excited states. In these highly excited levels, the hyperfine structure is negligible. The atomic state is therefore correctly specified by $i$) the principal quantum number $n$, \emph{ii}) the azimuthal quantum number $L$, \emph{iii}) the total angular momentum quantum number $J\in\left\llbracket\left|L-\frac{1}{2}\right|,\left|L+\frac{1}{2}\right|\right\rrbracket $
and \emph{iv}) the magnetic quantum number $M_{J}$ associated with
the projection of the total angular momentum onto the quantisation
axis of unit vector ${\bf e}_{q}$, i.e.  $\hat{J}_{q}\equiv\hat{{\bf J}}\cdot{\bf e}_{q}$.

In the nonretarded approximation, the interaction between
two Rydberg atoms near a medium can be described by the following effective Hamiltonian, including electric dipolar and quadrupolar contributions (see \cite{BS17} for details)
\begin{eqnarray}
\hat{H}_{\textrm{eff}} & = & \frac{1}{\epsilon_{0}}\hat{{\bf d}}_{A}\cdot{\bf \overline{T}}\left({\bf r}_{A},{\bf r}_{B}\right)\cdot\hat{{\bf d}}_{B}\nonumber \\
 & + & \frac{1}{\epsilon_{0}}\hat{{\bf d}}_{A}\cdot{\bf \overline{T}}\left({\bf r}_{A},{\bf r}_{B}\right)\otimes\nabla_{B}\bullet\hat{\overline{{\bf Q}}}_{B}\nonumber \\
 & + & \frac{1}{\epsilon_{0}}\hat{{\bf \overline{Q}}}_{A}\bullet\nabla_{A}\otimes{\bf \overline{T}}\left({\bf r}_{A},{\bf r}_{B}\right)\cdot\hat{{\bf d}}_{B}\nonumber \\
 & + & \frac{1}{\epsilon_{0}}\hat{{\bf \overline{Q}}}_{A}\bullet\nabla_{A}\otimes{\bf \overline{T}}\left({\bf r}_{A},{\bf r}_{B}\right)\otimes\nabla_{B}\bullet\hat{\overline{{\bf Q}}}_{B}\label{HamiltonienComplet-1}
\end{eqnarray}
where \emph{i}) $\hat{{\bf d}}_{K=A,B}=-e\hat{{\bf r}}_{K}$ and
$\hat{{\bf \overline{Q}}}_{K=A,B}=-\frac{e}{2}\hat{{\bf r}}_{K}\otimes\hat{{\bf r}}_{K}$
are the electric dipolar and quadrupolar moment operators, respectively,
of atom $K=A,B$, with $\hat{{\bf r}}_{K}$ denoting the position operator of the valence electron
in atom $K=A,B$ relative to the atomic centre of mass, \emph{ii}) ${\bf \overline{T}}\left({\bf r}_{A},{\bf r}_{B}\right)\equiv\lim_{\omega\rightarrow0^{+}}\left(\frac{\omega}{c}\right)^{2}{\bf \overline{G}}\left({\bf r}_{A},{\bf r}_{B},\omega\right)$, \emph{iii}) $\nabla_{K}$ is the gradient operator
with respect to the coordinates of atom $K=A,B$, and \emph{iv}) $\overline{a}\bullet\overline{b}\equiv\sum_{i,j}a_{ij}b_{ji}$
is the Frobenius product between two tensors $\overline{a}$
et $\overline{b}$ defined by their components $\left\{ a_{ij},b_{ij}\right\} $
in an orthonormal basis \cite{Buh12}.
As the dyadic Green's function it is derived from, the tensor ${\bf \overline{T}}$ comprises a free-space component,
$
{\bf \overline{T}}_{0}\left({\bf r}\right)=-\frac{1}{4\pi r_{AB}^{3}}\left({\bf \overline{I}}-3{\bf u}_{AB}\otimes{\bf u}_{AB}\right)
$, 
where $r_{AB}\equiv\left|{\bf r}_{A}-{\bf r}_{B}\right|$ and ${\bf u}_{AB}\equiv\frac{1}{r_{AB}}\left({\bf r}_{B}-{\bf r}_{A}\right)$, and a reflected part due to the presence of the fibre, denoted by ${\bf \overline{T}}_{1}$. The explicit form of ${\bf \overline{T}}_{1}$ is too cumbersome to be reproduced here, the expression of the reflected part of the dyadic Green's function, $\overline{\mathbf{G}}_1$, from which ${\bf \overline{T}}_{1}$ is deduced can be found in \cite{SLR20}.
Note that, in free-space, the dipole-dipole component in the first line in equation (\ref{HamiltonienComplet-1}) reduces to
\begin{eqnarray}
\hat{H}_{\textrm{eff},0} & = & -\frac{1}{4\pi\epsilon_{0}r_{AB}^{3}}\left[{\bf \hat{d}}_{A}\cdot{\bf \hat{d}}_{B}-3\left({\bf \hat{d}}_{A}\cdot{\bf u}_{AB}\right)\left({\bf \hat{d}}_{B}\cdot{\bf u}_{AB}\right)\right]\label{eq:HamiltonienElectrostatique-1}
\end{eqnarray}
which allows one to recover, to the second order of the perturbation
theory, the electrostatic potential between atoms $A$ and $B$ respectively prepared in states $\left( \left|m \right\rangle, \left|n \right\rangle \right)$, $U_{AB}^{\left(0\right)}\left({\bf r}_{A},{\bf r}_{B}\right)=-\frac{C_{6}}{r_{AB}^{6}}$, 
with
\begin{eqnarray}
C_{6} & =\frac{1}{16 \hbar \pi^{2}\epsilon_{0}^{2}}\sum_{k,l}\frac{\left|{\bf d}_{mk}^{A}\cdot{\bf d}_{nl}^{B}-3\left({\bf d}_{mk}^{A}\cdot{\bf u}_{AB}\right)\left({\bf d}_{nl}^{B}\cdot{\bf u}_{AB}\right)\right|^{2}}{\omega_{mk}^{A}+\omega_{nl}^{B}}\label{eq:ExpressionC6-1}
\end{eqnarray}
where $\left( \left|k \right\rangle, \left|l \right\rangle \right)$ denote intermediate states of atoms $A$ and $B$.

In the following sections, we study the interaction potential between two  $^{87}$Rb atoms, prepared in various Rydberg states, that we numerically obtained either through direct diagonalisation of the effective Hamiltonian, equation (\ref{HamiltonienComplet-1}), in a truncated basis or via second order perturbation theory. The truncated basis typically comprises states $\left\{ \left|n^{(A)} L^{(A)} J^{(A)} M^{(A)}_J ; n^{(B)} L^{(B)} J^{(B)} M^{(B)}_J \right\rangle  \right\}$ which are directly coupled by the Hamiltonian, equation (\ref{HamiltonienComplet-1}), to the two-atom state of interest $\left|n L J M_J ; n L J M_J\right\rangle$, with $n^{(A)},n^{(B)}$ ranging from $n_{\text{min}}$ to $n_{\text{max}}$ and $n_{\text{min}} \approx n-10$ and $n_{\text{max}} \approx n+10$. We check the convergence of the calculations by ensuring that adding (subtracting) 1 to $n_{\text{max}}$ (resp. $n_{\text{min}}$) does not significantly modify our results. 

\section{Interaction of two rubidium atoms in the state $\left|nS_{1/2}\right\rangle $\label{IntSS}}

In this section, we study the interaction between
atoms $\left(A,B\right)$, prepared
in the same Rydberg state $\left|nS_{1/2}\right\rangle $, for $n\geq30$.
We show how the presence of the nanofibre modifies the potential $U_{AB}$
in the so-called lateral configuration, i.e. when $R_{A}=R_{B}=R$ and
$\Delta\phi=0$, and for $n=30$ (Sec. \ref{DeltaZ}). Then, using a
simplified model, we qualitatively account for the behaviour observed (Sec.
\ref{subsec:Modele-simplifie}) and relate it
to the appearance of new couplings induced by a fibre-assisted symmetry
breaking (Sec. \ref{subsec:Apparition-de-nouveaux}). We study how
previous results evolve when the principal quantum number, $n$, varies
(Sec. \ref{NombreQuantiquePrincipal}). Finally, we briefly examine
other geometric configurations, $\Delta\phi\neq0$, in which the interatomic
axis is no longer parallel to the fibre axis, and which give rise to various behaviours for $U_{AB}$ (Sec. \ref{DependanceDeltaPhi}).  In Secs (\ref{DeltaZ}-\ref{NombreQuantiquePrincipal}), we restrict
ourselves to the lateral configuration $\Delta\phi=0$, $R_{A}=R_{B}=R$
and choose the quantisation axis along $\left(Oz\right)$. In Sec.
\ref{DependanceDeltaPhi}, we explore configurations for which $\Delta\phi\neq0$.

\subsection{Dependence on the lateral distance, $\Delta z$\label{DeltaZ}}

\subsubsection*{Numerical results}

\begin{figure*}
\begin{centering}
\includegraphics[width=17cm]{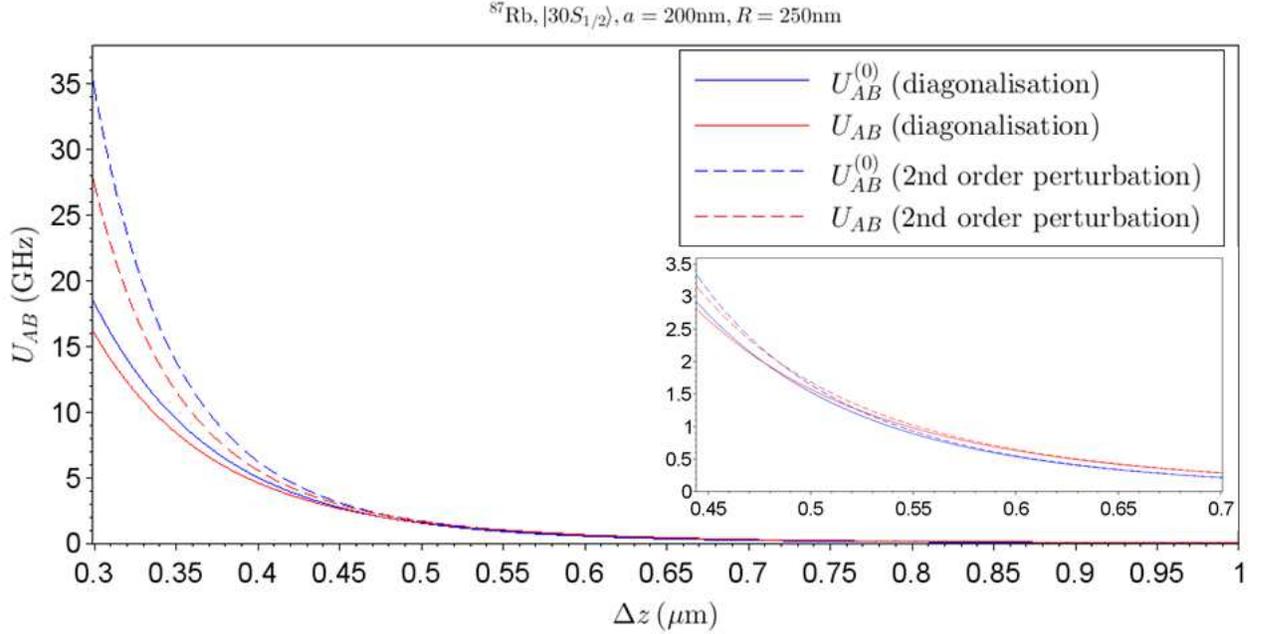}
\par\end{centering}
\caption{\textbf{Interaction between two $^{87}\mbox{Rb}$ atoms,
$\left(A,B\right)$, prepared in the state $|30S_{1/2}\rangle$ and
located in free-space or near an optical nanofibre. }Van
der Waals potentials in free-space, $U_{AB}^{\left(0\right)}$ (blue
curves), and near a nanofibre, $U_{AB}$ (red curves),
are plotted as functions of the lateral distance $\Delta z$ and calculated
either \emph{i}) through direct diagonalisation of the Hamiltonian,
equation (\ref{HamiltonienComplet-1}), including quadrupolar interactions
(full-line curves), or \emph{ii}) through second-order perturbation
theory relative to the same Hamiltonian (dashed-line curves). The
insert shows a zoom of the main figure in the range $0.45~\mu\mbox{m}\protect\leq\Delta z\protect\leq0.7~\mu\mbox{m}$.
The nanofibre radius is $a=200~\mathrm{nm}$, $\Delta\phi=0$, $R_{A}=R_{B}=R=250~\mathrm{nm}$
and the quantisation axis is along $\left(Oz\right)$.}

\label{FigPot30STotal}
\end{figure*}

In figure \ref{FigPot30STotal}, we show variations with $\Delta z$
of the potential $U_{AB}$ when atoms are prepared in the same state $\left|30S_{1/2}\right\rangle $
\footnote{Since the results we obtained do not depend on $M_{J}$, we merely
designate the atomic state by $\left|30S_{1/2}\right\rangle $ in
figure \ref{FigPot30STotal}.} and located either \emph{i}) in free-space (potential $U_{AB}^{\left(0\right)}$,
blue curves), or \emph{ii}) at a distance $R=250~\mbox{nm}$ from
the axis of a nanofibre of radius $a=200~\mathrm{nm}$ (potential
$U_{AB}$, red curves). This potential coincides with the energy shift
of the state $\left|30S_{1/2},M_{J}=\pm1/2\right\rangle \otimes\left|30S_{1/2},M_{J}=\pm1/2\right\rangle $
induced by the Hamiltonian equation (\ref{HamiltonienComplet-1}). This is calculated either a) through diagonalisation of the Hamiltonian
equation (\ref{HamiltonienComplet-1}) (full-line curves), or b) using
second-order perturbation theory relative to the same Hamiltonian
(dashed-line curves). In the considered range of distances, i.e. for
\emph{not too short} distances, the perturbation induced by the Hamiltonian equation (\ref{HamiltonienComplet-1}) on the initial state $\left|30S_{1/2},M_{J}=\pm1/2\right\rangle \otimes\left|30S_{1/2},M_{J}=\pm1/2\right\rangle $
remains moderate and it is possible to adiabatically follow its energy.

\begin{figure*}
\begin{centering}
\includegraphics[width=17cm]{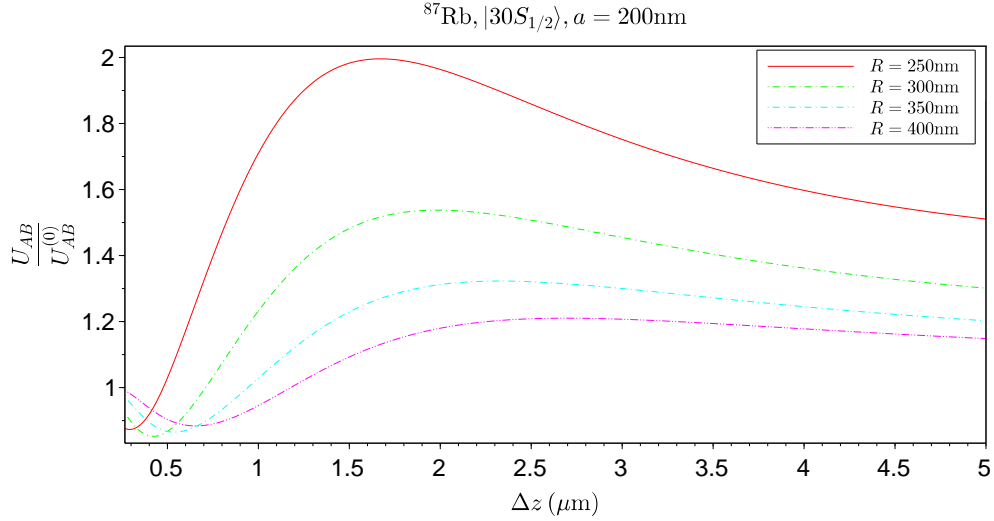}
\par\end{centering}
\caption{\textbf{Interaction between two $^{87}\mbox{Rb}$ atoms,
$\left(A,B\right)$, prepared in the state $|30S_{1/2}\rangle$ and
located in free-space or near an optical nanofibre.} We
consider the configuration $\Delta\phi=0$, $R_{A}=R_{B}=R$ and fix
the quantisation axis along $\left(Oz\right)$. The ratio $\left(\nicefrac{U_{AB}}{U_{AB}^{\left(0\right)}}\right)$
of the van der Waals interaction potentials between the atoms located
i) at a distance $R=\left(250,300,350,400\right)\mathrm{nm}$ from
an optical nanofibre of radius $a=200~\mathrm{nm}$ $\left(\mbox{numerator }U_{AB}\right)$,
and ii) in free-space $\left(\mbox{denominator }U_{AB}^{\left(0\right)}\right)$
is plotted as a function of the lateral distance $\Delta z$.}
\label{FigRapportVideFibre30S}
\end{figure*}

In figure \ref{FigRapportVideFibre30S}, we show the variations
with $\Delta z$ of the ratio $\left(\nicefrac{U_{AB}}{U_{AB}^{\left(0\right)}}\right)$
of the interaction potentials when atoms are located \emph{i}) at a distance $R=\left(250,300,350,400\right)\mbox{nm}$
from the axis of a nanofibre of radius $a=200$ nm (numerator $U_{AB}$) and \emph{ii})
in free-space (denominator $U_{AB}^{(0)}$). The results presented here were obtained through direct
diagonalisation of the Hamiltonian equation (\ref{HamiltonienComplet-1}).

\begin{figure*}
\begin{centering}
\includegraphics[width=16cm]{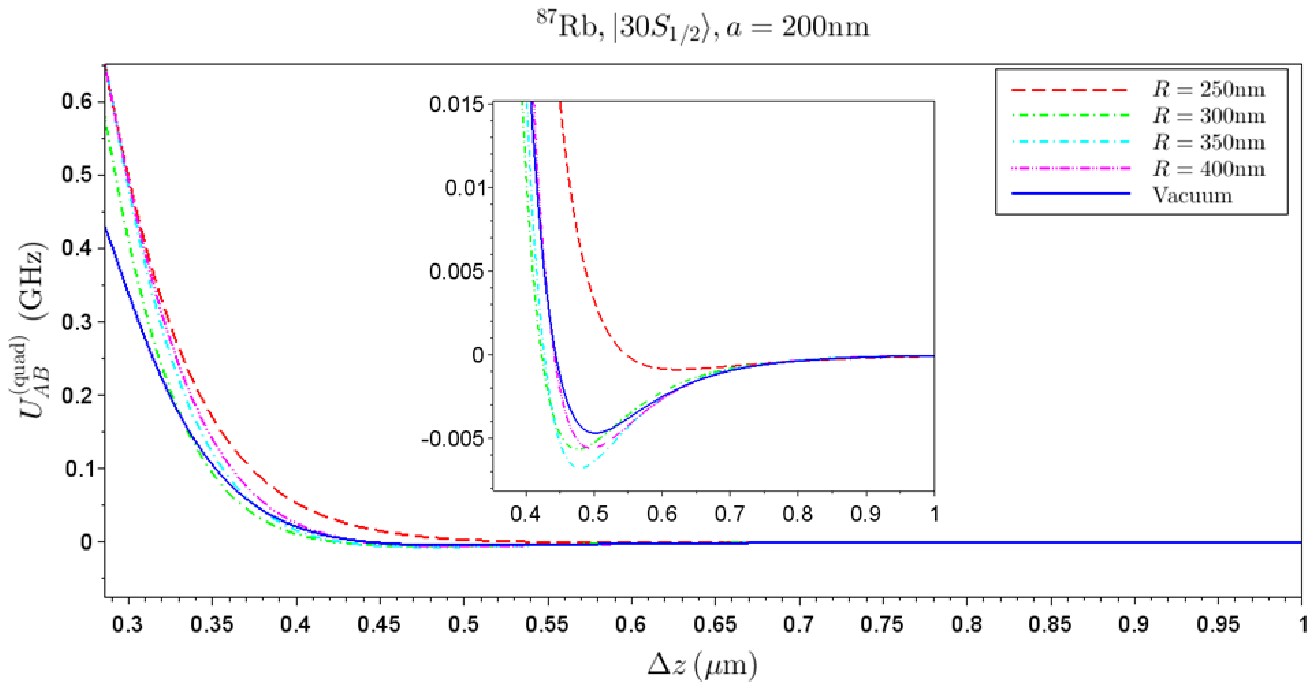}
\par\end{centering}
\caption{\textbf{Interaction between two $^{87}\mbox{Rb}$ atoms,
$\left(A,B\right)$, prepared in the state $|30S_{1/2}\rangle$ and
located in free-space or near an optical nanofibre: contribution
of quadrupolar transitions.} The quadrupolar contribution, \textbf{$U_{AB}^{\left(\mathrm{quad}\right)}$},
to the van der Waals potential in free-space (full-line blue curve)
and near an optical nanofibre (dashed-line curves) is
plotted as a function of $\Delta z$. The nanofibre radius is $a=200~\mathrm{nm}$,
$\Delta\phi=0$, $R_{A}=R_{B}=\left(250,300,350,400\right)\mathrm{nm}$
and the quantisation axis is fixed along $\left(Oz\right)$. The insert
shows a zoom in the vertical direction of the main figure in the range $0.4\mu\mbox{m}\protect\leq\Delta z\protect\leq1\mu\mbox{m}$.}
\label{FigPot30SQuad}
\end{figure*}

Finally, in figure \ref{FigPot30SQuad}, we show, as a function
of $\Delta z$, the quadrupolar contribution
to the interaction potential, \textbf{$U_{AB}^{\left(\mathrm{quad}\right)}$}, when atoms are located either \emph{i}) in free-space
(full-line blue curve), or \emph{ii}) at a distance $R_{A}=R_{B}=\left(250,300,350,400\right)\mathrm{nm}$
from a nanofibre of radius $a=200~\mathrm{nm}$ (dashed-line curves).

\subsubsection*{Analysis and comments}

The potential plotted in figure \ref{FigPot30STotal} is \emph{repulsive}
in free-space as well as in the presence of the nanofibre. As seen in
figure \ref{FigRapportVideFibre30S}, the potential is weaker (resp.
larger) in the presence of the nanofibre at short (resp. large) lateral
separations $\Delta z$, i.e. $\nicefrac{U_{AB}}{U_{AB}^{\left(0\right)}}<1$
(resp. $\nicefrac{U_{AB}}{U_{AB}^{\left(0\right)}}>1$). For example,
for $R=250~\mbox{nm}$ $\left(\mbox{resp. }R=400~\mbox{nm}\right)$, the
potential is enhanced for $\Delta z\gtrsim0.5~\mu\mbox{m}$ $\left(\mbox{resp. }\Delta z\gtrsim1.2~\mu\mbox{m}\right)$.

Figure \ref{FigPot30STotal} further shows that exact and perturbative
results coincide when atoms are sufficiently far apart from
each other, i.e. for distances $\Delta z$ larger than the van
der Waals radius\footnote{Van der Waals radius $R_{\mathrm{vdW}}\left(nS_{1/2}\right)$
is defined as the distance $\Delta z$ between two atoms at which
the approximation $U_{AB}^{\left(0\right)}\left(nS_{1/2}\right)=-\frac{C_{6}^{\left(0\right)}\left(nS_{1/2}\right)}{\Delta z^{6}}$
becomes valid.} $R_{\mbox{vdW}}\approx0.6$~$\mu\mbox{m}$.
In this perturbative regime, quadrupolar effects are negligible, as
can be seen in figure \ref{FigPot30SQuad}, and the interaction potential
is therefore dominated by the contribution of dipolar transitions,
both in free-space and near the nanofibre. In particular, $U_{AB}^{\left(0\right)}$
approximately follows the law $U_{AB}^{\left(0\right)}\approx-\frac{C_{6}^{\left(0\right)}\left(|30S_{1/2}\rangle\right)}{\Delta z^{6}}$,
with $C_{6}^{\left(0\right)}\left(|30S_{1/2}\rangle\right)\approx-26~\mathrm{MHz.\left(\mu m\right)^{6}}$.
Moreover, as shown in figure \ref{FigRapportVideFibre30S}, for $\Delta z\gg R_{\mathrm{vdW}}$,
the ratio $\left(\nicefrac{U_{AB}}{U_{AB}^{\left(0\right)}}\right)$
varies slowly as a function of $\Delta z$, and can be considered
locally constant, i.e. $\left(\nicefrac{U_{AB}}{U_{AB}^{\left(0\right)}}\right)\approx\alpha$,
with $\alpha\approx2$ around $\Delta z\approx1.6~\mu\mbox{m}$ for $R=250~\mbox{nm}$). Hence, the potential $U_{AB}$ locally follows the usual law $U_{AB}\approx-\frac{C_{6}\left(30S_{1/2}\right)}{\Delta z^{6}}$,
with $C_{6}\left(30S_{1/2}\right)=\alpha C_{6}^{\left(0\right)}\left(30S_{1/2}\right)$.
In other words, the presence of the nanofibre multiplies the $C_{6}$
coefficient by a factor $\alpha$ and, hence, the blockade radius
$r_{\mbox{blockade}}\propto\left(C_{6}\right)^{\frac{1}{6}}$ by a
factor $\alpha^{\frac{1}{6}}$. The ``constant'' $\alpha$ is larger
than $1$ and increases as atoms get closer to the nanofibre, i.e.
for ``small'' $R$'s. As we shall see in Sec. \ref{NombreQuantiquePrincipal},
$\alpha$ does not depend on the principal quantum number, $n$.

\subsection{Simplified model : $\pi-\pi$ coupling \label{subsec:Modele-simplifie}}

In this section, we develop a simplified model to qualitatively account
for the main features observed on the potential $U_{AB}$. 

We denote by $\left|n\right\rangle _{A}\left|n\right\rangle _{B}\equiv\left|30S_{\nicefrac{1}{2}}\right\rangle _{A}\left|30S_{\nicefrac{1}{2}}\right\rangle _{B}$
the state in which atoms $A$ and $B$ are initially prepared. The
partial contribution to the potential $U_{AB}$ due to the coupling
of $\left|n\right\rangle _{A}\left|n\right\rangle _{B}$ to another
state $\left|k\right\rangle _{A}\left|l\right\rangle _{B}$ by the
dipole-dipole interaction Hamiltonian is $U_{kl}=\frac{1}{\hbar\epsilon_{0}^{2}\Delta_{kl}}\left|{\bf d}_{nl}^{B}\cdot{\bf \overline{T}}\left({\bf r}_{B},{\bf r}_{A}\right)\cdot{\bf d}_{nk}^{A}\right|^{2}$,
with $\Delta_{kl}\equiv\omega_{nk}^{A}+\omega_{nl}^{B}$, and $\hbar\omega_{nk}\equiv E_{n}-E_{k}$
is the energy of the transition $\left|k\right\rangle \rightarrow\left|n\right\rangle $.
In this expression, the term 
\begin{eqnarray*}
\left|{\bf d}_{nl}^{B}\cdot{\bf \overline{T}}\left({\bf r}_{B},{\bf r}_{A}\right)\cdot{\bf d}_{nk}^{A}\right|^{2} & =\left({\bf d}_{nk}^{A}\cdot{\bf \overline{T}}\left({\bf r}_{A},{\bf r}_{B}\right)\cdot{\bf d}_{nl}^{B}\right)\left({\bf d}_{ln}^{B}\cdot{\bf \overline{T}}\left({\bf r}_{B},{\bf r}_{A}\right)\cdot{\bf d}_{kn}^{A}\right)
\end{eqnarray*}
can be interpreted as the exchange of two (real or virtual) photons
between the atomic dipoles ${\bf d}_{A}$ and ${\bf d}_{B}$ propagated
from $A$ to $B$ and from $B$ to $A$ by the functions ${\bf \overline{T}}\left({\bf r}_{B},{\bf r}_{A}\right)$
and ${\bf \overline{T}}\left({\bf r}_{A},{\bf r}_{B}\right)$, respectively.
In the nonretarded approximation, this propagation is considered instantaneous.
Since ${\bf T}={\bf T}_{0}+{\bf T}_{1}$, one gets 
\begin{eqnarray}
U_{kl} & = & U_{kl}^{\left(0\right)}+U_{kl}^{\left(\mathrm{vac-fib}\right)}+U_{kl}^{\left(\mathrm{fib-fib}\right)}\label{eq:DecompositionVdW}\\
U_{kl}^{\left(0\right)} & = & \frac{1}{\hbar\epsilon_{0}^{2}\Delta_{kl}}\left|{\bf d}_{nl}^{B}\cdot{\bf \overline{T}}_{0}\left({\bf r}_{B},{\bf r}_{A}\right)\cdot{\bf d}_{nk}^{A}\right|^{2}\label{eq:PotVac}\\
U_{kl}^{\left(\mathrm{vac-fib}\right)} & = & \frac{2}{\hbar\epsilon_{0}^{2}\Delta_{kl}}\mathrm{Re}\left[\left({\bf d}_{nl}^{B}\cdot{\bf \overline{T}}_{0}\left({\bf r}_{B},{\bf r}_{A}\right)\cdot{\bf d}_{nk}^{A}\right)\right.\label{eq:PotVacFib}\\
 &  & \times \left.\left({\bf d}_{kn}^{A}\cdot{\bf \overline{T}}_{1}\left({\bf r}_{B},{\bf r}_{A}\right)\cdot{\bf d}_{ln}^{B}\right)\right]\nonumber \\
U_{kl}^{\left(\mathrm{fib-fib}\right)} & = & \frac{1}{\hbar\epsilon_{0}^{2}\Delta_{kl}}\left|{\bf d}_{nl}^{B}\cdot{\bf \overline{T}}_{1}\left({\bf r}_{B},{\bf r}_{A}\right)\cdot{\bf d}_{nk}^{A}\right|^{2}\label{eq:PotFib}
\end{eqnarray}
In this formula, $U_{kl}^{\left(0\right)}$ can be associated with
the direct exchange of two photons in free-space, $U_{kl}^{\left(\mathrm{vac-fib}\right)}$
with the exchange of one photon via free-space and one photon
via reflection onto the nanofibre, $U_{kl}^{\left(\mathrm{fib-fib}\right)}$
with the exchange of two photons via reflection onto the nanofibre.
Moreover, with these notations, we have
\begin{eqnarray}
\frac{U_{AB}}{U_{AB}^{\left(0\right)}} & = & 1+\frac{\sum_{kl}U_{kl}^{\left(\mathrm{vac-fib}\right)}}{\sum_{kl}U_{kl}^{\left(0\right)}}+\frac{\sum_{kl}U_{kl}^{\left(\mathrm{fib-fib}\right)}}{\sum_{kl}U_{kl}^{\left(0\right)}}\label{eq:RapportFibreVide}
\end{eqnarray}

\subsubsection*{Simplified model}

\begin{figure*}
\begin{centering}
\includegraphics[width=16cm]{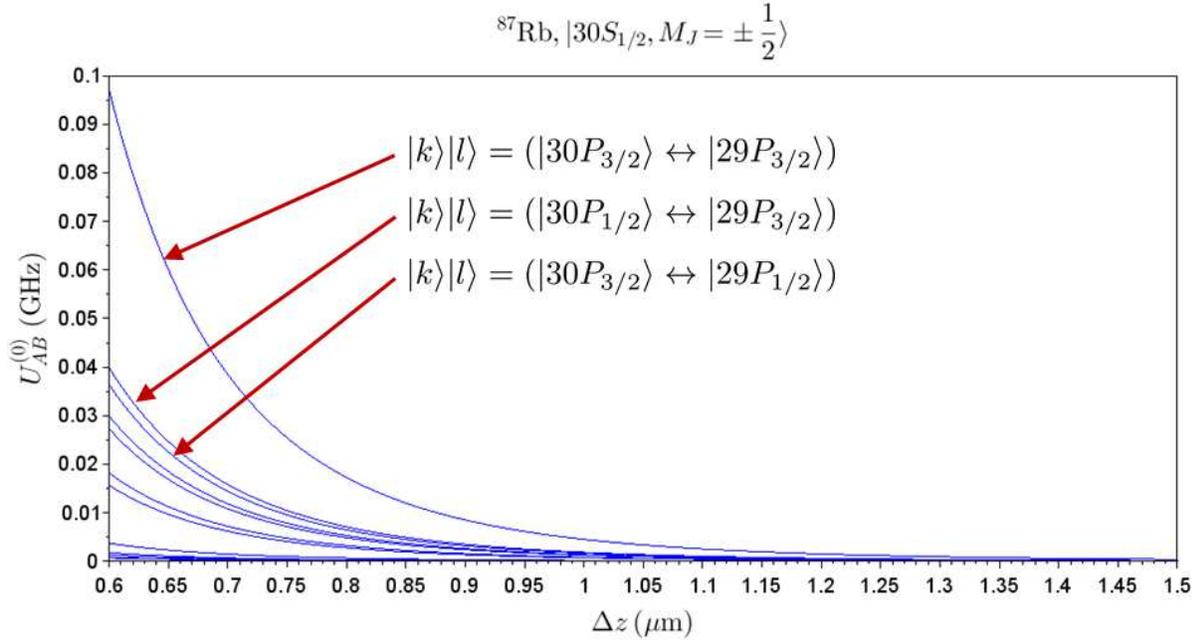}
\par\end{centering}
\caption{\textbf{Interaction between two $^{87}\mbox{Rb}$ atoms,
$\left(A,B\right)$, prepared in the state $|30S_{1/2}\rangle$ and
located in free-space : partial contributions, $U_{kl}^{\left(0\right)}$,
of the main couplings $\left|n\right\rangle _{A}\left|n\right\rangle _{B}\leftrightarrow\left|k\right\rangle _{A}\left|l\right\rangle _{B}$.}
The partial contributions are plotted as functions of the lateral
distance, $\Delta z$, between the two atoms. We fixed $\Delta\phi=0$,
$R_{A}=R_{B}=R$ and the quantisation axis was chosen along $\left(Oz\right)$.}

\label{FigDetailPertubation30S}
\end{figure*}

We start by a few remarks on the interaction potential in free-space.
The numerator $\left|{\bf d}_{nl}^{B}\cdot{\bf \overline{T}}_{0}\left({\bf r}_{B},{\bf r}_{A}\right)\cdot{\bf d}_{nk}^{A}\right|^{2}$
of the partial contribution $U_{kl}^{\left(0\right)}$ (see equation \ref{eq:PotVac}) is always positive,
contrary to the denominator $\Delta_{kl}$ : the repulsive or attractive
nature of the total potential in free-space, $U_{AB}^{\left(0\right)}=\sum_{kl}U_{kl}^{\left(0\right)}$,
is therefore determined by the sign of the denominators $\Delta_{kl}$ and
the relative magnitudes of the dipole momenta of each transition. In figure
\ref{FigDetailPertubation30S} are plotted the main contributions,
$U_{kl}^{\left(0\right)}$, to the total potential in free-space, $U_{AB}^{\left(0\right)}$,
due to the couplings $\left|n\right\rangle _{A}\left|n\right\rangle _{B}\leftrightarrow\left|k\right\rangle _{A}\left|l\right\rangle _{B}$.

The main contributions both coincide with the highest branch
and are due to the coupling of $\left|n\right\rangle _{A}\left|n\right\rangle _{B}$
with the states $\left|k\right\rangle _{A}\left|l\right\rangle _{B}$
and $\left|l\right\rangle _{A}\left|k\right\rangle _{B}$, briefly
denoted by $\left(\left|k\right\rangle _{A}\leftrightarrow\left|l\right\rangle _{B}\right)$,
with $\left|k\right\rangle \equiv\left|30P_{3/2},M_{j}=\pm\frac{1}{2}\right\rangle $
and $\left|l\right\rangle \equiv\left|29P_{3/2},M_{j}=\pm\frac{1}{2}\right\rangle $.
This coupling is of ``$\pi-\pi$'' type, i.e. in this coupling scheme,
each atom undergoes a $\pi$ transition along which the magnetic quantum
number $M_{j}$ remains unchanged.

The lower two branches are also associated with $\pi-\pi$-type
couplings, i.e. $\left|n\right\rangle _{A}\left|n\right\rangle _{B}\rightarrow\left(\left|k\right\rangle _{A}\leftrightarrow\left|l\right\rangle _{B}\right)$,
respectively with 
\begin{eqnarray*}
i) & \left\{ \left|k\right\rangle \equiv\left|30P_{\nicefrac{1}{2}}\right\rangle ,\left|l\right\rangle \equiv\left|29P_{\nicefrac{3}{2}}\right\rangle \right\} \\
ii) & \left\{ \left|k\right\rangle \equiv\left|30P_{\nicefrac{3}{2}}\right\rangle ,\left|l\right\rangle \equiv\left|29P_{\nicefrac{1}{2}}\right\rangle \right\} 
\end{eqnarray*}

In our simplified model, we suppose that the potential in free-space,
$U_{AB}^{\left(0\right)}$, is solely determined by the coupling $\left|n\right\rangle _{A}\left|n\right\rangle _{B}\rightarrow\left(\left|k\right\rangle _{A}\leftrightarrow\left|l\right\rangle _{B}\right)$,
with $\left|k\right\rangle \equiv\left|30P_{3/2},M_{j}=\pm\frac{1}{2}\right\rangle $
and $\left|l\right\rangle \equiv\left|29P_{3/2},M_{j}=\pm\frac{1}{2}\right\rangle $,
i.e. $U_{AB}^{\left(0\right)}\approx U_{kl}^{\left(0\right)}$, and
so is the potential in the presence of the nanofibre, $U_{AB}$, i.e. $U_{AB}\approx U_{kl}$. 
The dipoles $\left({\bf d}_{nk}^{A},{\bf d}_{nl}^{B}\right)$ associated
with the main coupling, of $\pi-\pi$ type, are real and along the
quantisation axis, $\left(Oz\right)$, i.e. ${\bf d}_{nk}^{A}=d_{0,nk}^{A}{\bf e}_{z}$
and ${\bf d}_{nl}^{B}=d_{0,nl}^{B}{\bf e}_{z}$. The ratio $\left(\nicefrac{U_{AB}}{U_{AB}^{\left(0\right)}}\right)$ in
equation (\ref{eq:RapportFibreVide}) therefore takes the simple form

\begin{equation}
\frac{U_{AB}}{U_{AB}^{\left(0\right)}}=\left(1+2\pi\left(\Delta z\right)^{3}\left[{\bf \overline{T}}_{1}\left({\bf r}_{B},{\bf r}_{A}\right)\right]_{zz}\right)^{2}\label{eq:RapportFibreVac1Trans}
\end{equation}
where we used $\left[{\bf \overline{T}}_{0}\left({\bf r}_{B},{\bf r}_{A}\right)\right]_{zz}=\frac{1}{2\pi\left(\Delta z\right)^{3}}$
and the reality of the function $\left[{\bf \overline{T}}_{1}\left({\bf r}_{B},{\bf r}_{A}\right)\right]_{zz}$.
Remarkably, this ratio does not depend on dipoles $d_{0,nk}^{A}$
and $d_{0,nl}^{B}$, and the decrease/enhancement of the interaction
potential induced by the introduction of the fibre with respect to
the free-space is only determined by the sign of $\left[{\bf \overline{T}}_{1}\left({\bf r}_{B},{\bf r}_{A}\right)\right]_{zz}$.

\subsubsection*{Half-space approximation}

The function $\left[{\bf \overline{T}}_{1}\left({\bf r}_{B},{\bf r}_{A}\right)\right]_{zz}^{\left(\mathrm{fibre}\right)}$
can only be numerically computed from the expression of the reflected dyadic Green's function, $\overline{\mathbf{G}}_1$ which can be found in \cite{SLR20}. If, however, $R$ and $\Delta z$
are short ``enough'', the fibre surface can be regarded as a plane
of Cartesian equation $x=a$, and $\left[{\bf \overline{T}}_{1}\left({\bf r}_{B},{\bf r}_{A}\right)\right]_{zz}^{\left(\mathrm{fibre}\right)}$
approximately coincides with the function $\left[{\bf \overline{T}}_{1}\left({\bf r}_{B},{\bf r}_{A}\right)\right]_{zz}^{\left(\mathrm{plane}\right)}$
associated with the dielectric half-space $\left(x<a\right)$ the expression of which can be found, e.g., in \cite{Buh12}
\begin{eqnarray*}
\left[{\bf \overline{T}}_{1}\left({\bf r}_{B},{\bf r}_{A}\right)\right]_{zz}^{\left(\mathrm{fibre}\right)} \approx\left[{\bf \overline{T}}_{1}\left({\bf r}_{B},{\bf r}_{A}\right)\right]_{zz}^{\left(\mathrm{plane}\right)} =\frac{1}{4\pi}\left(\frac{n\left(0\right)^{2}-1}{n\left(0\right)^{2}+1}\right)\frac{\left(2X\right)^{2}-2\left(\Delta z\right)^{2}}{\left[\left(2X\right)^{2}+\left(\Delta z\right)^{2}\right]^{\frac{5}{2}}}
\end{eqnarray*}
where $X\equiv R-a$ is the distance of atoms $A$ and $B$ to the
fibre surface.

\begin{figure*}
\begin{centering}
\includegraphics[width=16cm]{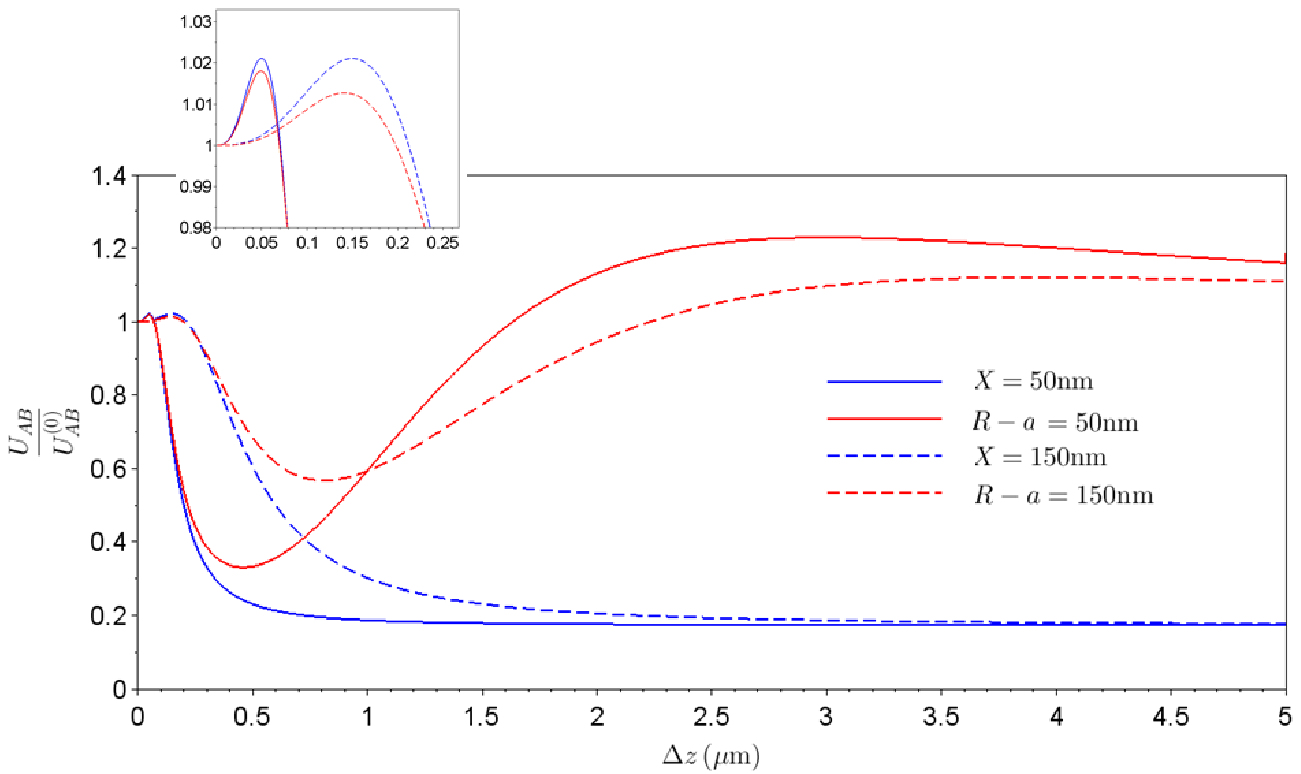}
\par\end{centering}
\caption{\textbf{Interaction between two $^{87}\mbox{Rb}$ atoms,
$\left(A,B\right)$, prepared in the state $|30S_{1/2}\rangle$ in
the neighbourhood of a nanofibre and a dielectric half-space.} We consider
the configuration $\Delta\phi=0$, $R_{A}=R_{B}=R$ and choose the
quantisation axis along $\left(Oz\right)$. We represented as functions
of the lateral distance $\Delta z$ the ratios $\left(\nicefrac{U_{AB}}{U_{AB}^{\left(0\right)}}\right)$
of van der Waals interaction potentials between two atoms located
i) at the distance $X=R-a=\left(50,150\right)\mbox{nm}$ from the
surface of a nanofibre of radius $a=200~\mathrm{nm}$ (red curves)
or a dielectric half-space of same index (blue curves), $\left(\mbox{numerator }U_{AB}\right)$,
and ii) in free-space $\left(\mbox{denominator }U_{AB}^{\left(0\right)}\right)$.}

\label{FigRapportFibreVidePiPi}
\end{figure*}

In figure \ref{FigRapportFibreVidePiPi} is plotted, as a function of
the distance $\Delta z$, the ratio $\left(\nicefrac{U_{AB}}{U_{AB}^{\left(0\right)}}\right)$
of i) the potentials between the two atoms located at the distance $X=R-a=\left(50,150\right)\mbox{nm}$
from the surface of a nanofibre of radius $a=200~\mathrm{nm}$ (red
curves) or a dielectric half-space of same optical index (blue curves),
$\left(\mbox{numerator }U_{AB}\right)$, and ii) in free-space $\left(\mbox{denominator }U_{AB}^{\left(0\right)}\right)$.
The results presented in figure \ref{FigRapportFibreVidePiPi} were
obtained in the framework of our simplified model. When $\Delta z\ll X,\left(R-a\right)$,
atoms do not ``see'' the nanofibre or dielectric half-space and
the direct exchange of photons dominates, i.e.
$\left(\nicefrac{U_{AB}}{U_{AB}^{\left(0\right)}}\right)\rightarrow1$.
As long as $\Delta z<a$, the results obtained with the half-space
and fibre coincide. When $\Delta z>a$, the half-space approximation
is no longer valid : in the fibre case, $U_{AB}$ first increases
with $\Delta z\left(>a\right)$, exceeds $U_{AB}^{\left(0\right)}$,
and reaches a maximum before slowly decreasing. Note that the maximum
is higher for atoms closer to the fibre. Unfortunately, because of
numerical issues appearing for large $\Delta z$, we were not yet
able to determine whether $\nicefrac{U_{AB}}{U_{AB}^{\left(0\right)}}$
tends towards a nonvanishing limiting value when $\Delta z\rightarrow+\infty$,
as figure \ref{FigRapportFibreVidePiPi} suggests.

\subsubsection*{Comparison with the full calculation}

Our simplified model, the results of which are presented in figure
\ref{FigRapportFibreVidePiPi}, qualitatively account for the actual
behaviour of the interaction potential, plotted in figure \ref{FigRapportVideFibre30S}
: \emph{i}) at short distance, $\Delta z<\left(\Delta z\right)_{\mathrm{lim}}$,
the presence of the fibre decreases the potential then $ii$) enhances
it for $\Delta z>\left(\Delta z\right)_{\mathrm{lim}}$, the value
$\left(\Delta z\right)_{\mathrm{lim}}$ increases with $\left(R-a\right)$
; finally iii) when $\Delta z\gg\left(R-a\right)$, the ratio $\left(\nicefrac{U_{AB}}{U_{AB}^{\left(0\right)}}\right)$
seems to tend towards a finite limit which is higher for lower values
of $\left(R-a\right)$. We underline, however, that our simplified
model severely underestimates the potential at short distances $\Delta z$,
since it neglects the contributions of couplings involving $\sigma^{\pm}$-type
transitions which enhance the potential with respect to free-space.

\subsection{Breaking of the rotation symmetry around the interatomic axis
and appearance of new couplings\label{subsec:Apparition-de-nouveaux}}

In the previous section, we qualitatively reproduced the main features
of the interaction potential in the presence of an optical nanofibre,
thanks to a simplified model restricted to the dominating $\pi-\pi$-type coupling. Quantitative discrepancies with the full treatment,
however, exist that we related to the existence of other couplings.
More precisely, there exist $\pi-\pi$-, $\pi-\sigma^{\pm}$-, $\sigma^{\pm}-\sigma^{\pm}$-,
and $\sigma^{\pm}-\sigma^{\mp}$-type couplings. We recall that the dipole
of a $\sigma^{\pm}$ transition writes ${\bf d}_{\pm}=\frac{d_{\pm}}{\sqrt{2}}\left({\bf e}_{x}\pm\mathrm{i}{\bf e}_{y}\right)$,
the dipole of a $\pi$ transition is ${\bf d}_{0}=d_{0}{\bf e}_{z}$,
and the partial contribution to the van der Waals potential
of the $\left|n\right\rangle _{A}\left|n\right\rangle _{B}\rightarrow\left|k\right\rangle _{A}\left|l\right\rangle _{B}$
coupling, denoted by $U_{kl}$, is proportional to $\left|{\bf d}_{nk}^{A}\cdot{\bf \overline{T}}\cdot{\bf d}_{nl}^{B}\right|^{2}$. 

In the considered configuration, the free-space propagator takes the following diagonal form in the basis $\left[{\bf e}_{i}\otimes{\bf e}_{j}\right]_{i,j=x,y,z}$
\begin{equation}
\overline{{\bf T}}_{0}\left({\bf r}_{A},{\bf r}_{B}\right)=\frac{1}{4\pi\left(\Delta z\right)^{3}}\left(\begin{array}{ccc}
-1 & 0 & 0\\
0 & -1 & 0\\
0 & 0 & 2
\end{array}\right)\label{eq:GreenVideConfigLateral}
\end{equation}
The rotation symmetry  around the interatomic axis of $\overline{\mathbf{T}}_0$ implies : i) $\left[{\bf \overline{T}}\right]_{xz}=\left[{\bf \overline{T}}\right]_{yz}=0$
hence $\left|{\bf d}_{0}\cdot\overline{{\bf T}}_{0}\left({\bf r}_{A},{\bf r}_{B}\right)\cdot{\bf d}_{\pm}\right|=0$,
and the $\pi-\sigma$-type couplings do not contribute to the potential
$U_{AB}^{\left(0\right)}$ ; ii) $\left[{\bf \overline{T}}\right]_{xx}=\left[{\bf \overline{T}}\right]_{yy}$
and $\left[{\bf \overline{T}}\right]_{yx}=0$ hence $\left|{\bf d}_{\pm}\cdot\overline{{\bf T}}_{0}\left({\bf r}_{A},{\bf r}_{B}\right)\cdot{\bf d}_{\pm}\right|=0$,
and $\sigma^{\pm}-\sigma^{\pm}$-type couplings do not contribute
to the potential $U_{AB}^{\left(0\right)}$.

The presence of a dielectric medium, fibre or half-space, breaks this
symmetry and some couplings, which were forbidden in free-space become allowed. In the considered so-called lateral configuration, it can be proved that ${\bf \overline{T}}_{1}$ takes the following generic form
\begin{equation}
{\bf \overline{T}}_{1}=\left(\begin{array}{ccc}
T_{xx} & 0 & T_{xz}\\
0 & T_{yy} & 0\\
-T_{xz} & 0 & T_{zz}
\end{array}\right)\label{eq:GreenFibConfigLateral}
\end{equation}
for both the fibre and half-space, with $T_{xx}\ne T_{yy}$ and $T_{xz}\ne0$, in general. We then check that $\left|{\bf d}_{0}\cdot\overline{{\bf T}}_{1}\left({\bf r}_{A},{\bf r}_{B}\right)\cdot{\bf d}_{\pm}\right|=\left|{\bf d}_{\pm}\cdot\overline{{\bf T}}_{1}\left({\bf r}_{A},{\bf r}_{B}\right)\cdot{\bf d}_{\pm}\right|\ne0$.

\begin{figure*}
\begin{centering}
\includegraphics[width=13cm]{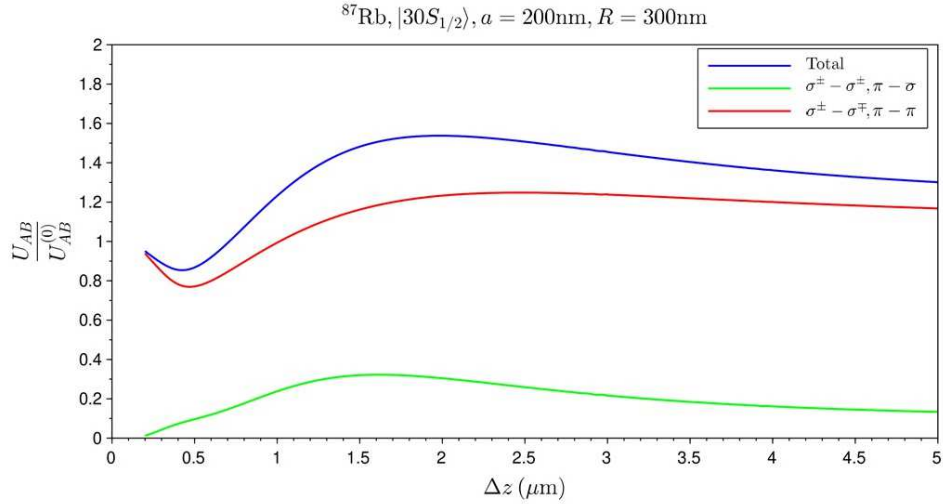}
\par\end{centering}
\caption{\textbf{Interaction between two $^{87}\mbox{Rb}$ atoms,
$\left(A,B\right)$, prepared in the state $\left|30S_{\nicefrac{1}{2}}\right\rangle $
in the neighbourhood of a nanofibre : partial contributions of the
couplings ``allowed'' and ``forbidden'' in free-space.} We fix $\Delta\phi=0$,
$R_{A}=R_{B}=R$ and choose the quantisation axis along $\left(Oz\right)$.
We represent the ratios $\left(\nicefrac{U_{AB}^{\left(\pi-\sigma\right)}+U_{AB}^{\left(\sigma^{\pm}-\sigma^{\pm}\right)}}{U_{AB}^{\left(0\right)}}\right)$
(green curve), $\left(\nicefrac{U_{AB}^{\left(\pi-\pi\right)}+U_{AB}^{\left(\sigma^{\pm}-\sigma^{\mp}\right)}}{U_{AB}^{\left(0\right)}}\right)$
(red curve), and $\frac{U_{AB}}{U_{AB}^{\left(0\right)}}$ (blue curve)
as functions of $\Delta z$ for $R=300$nm, where $U_{AB}^{\left(0\right)}$
designates the van der Waals potential between $A$ and $B$
in free-space, $U_{AB}^{\left(\gamma\right)}$ the partial contribution
of a coupling of type $\gamma=\pi-\pi,\pi-\sigma^{\pm},\sigma^{\pm}-\sigma^{\pm},\sigma^{\pm}-\sigma^{\mp}$
to the total van der Waals potential, $U_{AB}$, in the neighbourhood
of the fibre.}

\label{FigDetailContribution30S}
\end{figure*}

The partial contributions to the potential, $U_{AB}$, of the $\pi-\pi$,
$\pi-\sigma^{\pm}$, $\sigma^{\pm}-\sigma^{\pm}$ and $\sigma^{\pm}-\sigma^{\mp}$
couplings, respectively denoted by $U_{AB}^{\left(\pi-\pi\right)}$,
$U_{AB}^{\left(\pi-\sigma^{\pm}\right)}$, $U_{AB}^{\left(\sigma^{\pm}-\sigma^{\pm}\right)}$
and $U_{AB}^{\left(\sigma^{\pm}-\sigma^{\mp}\right)}$ can be calculated
through \emph{i}) a perturbative approach by restricting the couplings
to the relevant states $\left|k\right\rangle _{A}\left|l\right\rangle _{B}$,
or \emph{ii}) direct diagonalisation of the effective Hamiltonianen
by setting to zero the terms ${\bf d}_{nk}^{A}\cdot{\bf \overline{T}}\cdot{\bf d}_{ml}^{B}$
which correspond to unwanted transitions. In figure \ref{FigDetailContribution30S}
are plotted the ratios $\nicefrac{U_{AB}^{\left(\pi-\pi\right)}+U_{AB}^{\left(\sigma^{\pm}-\sigma^{\pm}\right)}}{U_{AB}^{\left(0\right)}}$
(green curve) and $\nicefrac{U_{AB}^{\left(\pi-\sigma\right)}+U_{AB}^{\left(\sigma^{\mp}-\sigma^{\pm}\right)}}{U_{AB}^{\left(0\right)}}$
(red curve) as functions of the distance $\Delta z$ and for $R=300\mathrm{nm}$.
These two ratios characterize the respective weights of the contributions
to the potential near the fibre due to couplings which
are allowed and forbidden in free-space. When atoms are very
close, i.e. when $\Delta z\rightarrow0$, the direct exchange of photons between atoms dominates and hence the weight of couplings forbidden in free-space is strongly decreased. The new contributions allowed by the fibre
at larger distances reinforce the enhancement of the potential : these
new couplings are responsible for the discrepancies between the results
obtained via the full calculation of the potential $U_{AB}$ (figure
\ref{FigRapportVideFibre30S}) and our simplified model involving
a single $\pi-\pi$ coupling (figure \ref{FigRapportFibreVidePiPi}).
We note, however, that this discrepancy remains moderate. This effect
shall be more dramatic with atoms prepared in a $P$ state, as we shall
see in Sec. \ref{IntPP}.

\subsection{Dependence on the principal quantum number, $n$\label{NombreQuantiquePrincipal}}

\begin{figure*}
\begin{centering}
\includegraphics[width=17cm]{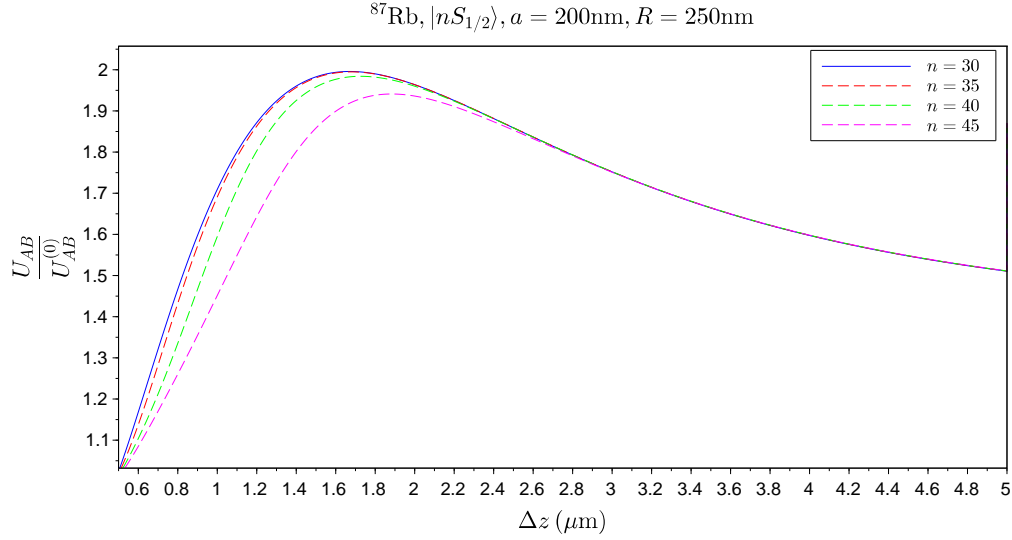}
\par\end{centering}
\caption{\textbf{Interaction between two  $^{87}\mbox{Rb}$ atoms,
$\left(A,B\right)$, prepared in the state $\left|nS_{\nicefrac{1}{2}}\right\rangle $
in the neighbourhood of a nanofibre : influence of the principal quantum
number, $n$.} We fix $\Delta\phi=0$, $R_{A}=R_{B}=R$ and choose
the quantisation axis along $\left(Oz\right)$. We represent as functions
of the lateral distance $\Delta z$ the ratio $\left(\nicefrac{U_{AB}}{U_{AB}^{\left(0\right)}}\right)$
of the van der Waals potentials between two atoms prepared in the state $\left|nS_{\nicefrac{1}{2}}\right\rangle $, for $n=30,35,40,45$,
and located \emph{i}) at the distance $R=250\mbox{nm}$ from the axis
of a nanofibre of radius $a=200\mathrm{nm}$ $\left(\mbox{numerator }U_{AB}\right)$
and \emph{ii}) in free-space $\left(\mbox{denominator }U_{AB}^{\left(0\right)}\right)$.}

\label{FigDepNiv}
\end{figure*}

Using the simplified model restricted to a single $\pi-\pi$ coupling
presented in Sec. \ref{subsec:Modele-simplifie},
we showed that the ratio $\left(\nicefrac{U_{AB}}{U_{AB}^{\left(0\right)}}\right)$
depends neither on the dipoles nor on the principal quantum number,
$n$. In the validity range of this model, the curves in figure \ref{FigRapportVideFibre30S}
are therefore universal, in the sense that they remain unchanged as
$n$ varies. To check this property we plotted in figure \ref{FigDepNiv}
the ratio $\left(\nicefrac{U_{AB}}{U_{AB}^{\left(0\right)}}\right)$
of the potentials when atoms are prepared in the same state $\left|nS_{\nicefrac{1}{2}}\right\rangle $
for $n=30,35,40,45$, and \emph{i}) located at the distance $R=250$nm
from the nanofibre $\left(\mbox{numerator }U_{AB}\right)$ and \emph{ii})
in free-space $\left(\mbox{denominator }U_{AB}^{\left(0\right)}\right)$
as a function of the distance $\Delta z$. Table \ref{TabDonn=0000E9ePotVacNivS}
gives the numerical values of $C_{6}^{\left(0\right)}$ coefficients
and van der Waals radius, $R_{\mathrm{vdW}}$, which characterize the interaction in free-space between two atoms prepared
in the same state $|nS_{1/2}\rangle$.

\begin{table}
\begin{centering}
\begin{tabular}{|c|c|c|c|c|}
\hline 
$n$ & 30 & 35 & 40 & 45\tabularnewline
\hline 
$C_{6}^{\left(0\right)}$ ($\mathrm{GHz.\left(\mu m\right)^{6}}$) & -0.026 & -0.185 & -0.98 & -4.23\tabularnewline
\hline 
$R_{\mathrm{VdW}}$($\mathrm{\mu m}$) & 0.5 & 0.6 & 0.9 & 1.5\tabularnewline
\hline 
\end{tabular}
\par\end{centering}
\caption{Numerical values of $C_{6}^{\left(0\right)}$ coefficients and van
der Waals radii, $R_{\mathrm{vdW}}$, which characterize the interaction
in free-space between two atoms prepared in the same state $|nS_{1/2}\rangle$,
for $n=30,35,40,45$.}

\label{TabDonn=0000E9ePotVacNivS}
\end{table}

The invariance of the ratio $\left(\nicefrac{U_{AB}}{U_{AB}^{\left(0\right)}}\right)$
with respect to $n$ is indeed observed
for $\Delta z\gtrsim2R_{\mathrm{vdW}}$, i.e. in the range where perturbation
theory is valid and where $U_{AB}^{\left(0\right)}$ scales as $\nicefrac{1}{\Delta z^{6}}$.
For $\Delta z\lesssim2R_{\mathrm{vdW}}$, the curves for different
$n$'s no longer coincide, though their shapes are much alike. Moreover,
as already noted in Sec. \ref{DeltaZ}, for $\Delta z\gtrsim2R_{\mathrm{vdW}}$,
the ratio $\left(\nicefrac{U_{AB}}{U_{AB}^{\left(0\right)}}\right)$
varies slowly -- especially for large $\Delta z$'s -- and may therefore
be considered locally constant. For any $n$, one has locally $\left(\nicefrac{U_{AB}}{U_{AB}^{\left(0\right)}}\right)\approx\alpha$,
i.e. $U_{AB}\approx-\frac{C_{6}\left(nS_{\nicefrac{1}{2}}\right)}{\Delta z^{6}}$
with $C_{6}\left(nS_{\nicefrac{1}{2}}\right)\approx\alpha C_{6}^{\left(0\right)}\left(nS_{\nicefrac{1}{2}}\right)$.
Introducing the nanofibre hence multiplies the $C_{6}$ coefficient
by the factor $\alpha$ -- and therefore the blockade radius $r_{\mbox{blockade}}\propto\left(C_{6}\right)^{\frac{1}{6}}$
by the factor $\alpha^{\frac{1}{6}}$. figure \ref{FigDepNiv} moreover
shows that this factor does not depend on the principal quantum number
-- it, however, depends on the distance of atoms to the fibre.

\subsection{Dependence on $\Delta\phi$\label{DependanceDeltaPhi}}

Until now, we focussed on the so-called lateral configuration defined by $\Delta\phi=0$
and $R_{A}=R_{B}=R$. In this section, we briefly investigate how $U_{AB}$ varies with $\Delta z$ for $\Delta\phi\neq0$,
keeping $R_{A}=R_{B}=R$ and choosing the quantisation axis along
$\left(Oz\right)$. The half-space approximation is, \emph{a priori}, no longer
applicable for this new type of configuration nor is the simplified
model restricted to a single $\pi-\pi$ coupling because the interatomic
axis does not coincide with the quantisation axis.

\begin{figure*}
\begin{centering}
\includegraphics[width=16cm]{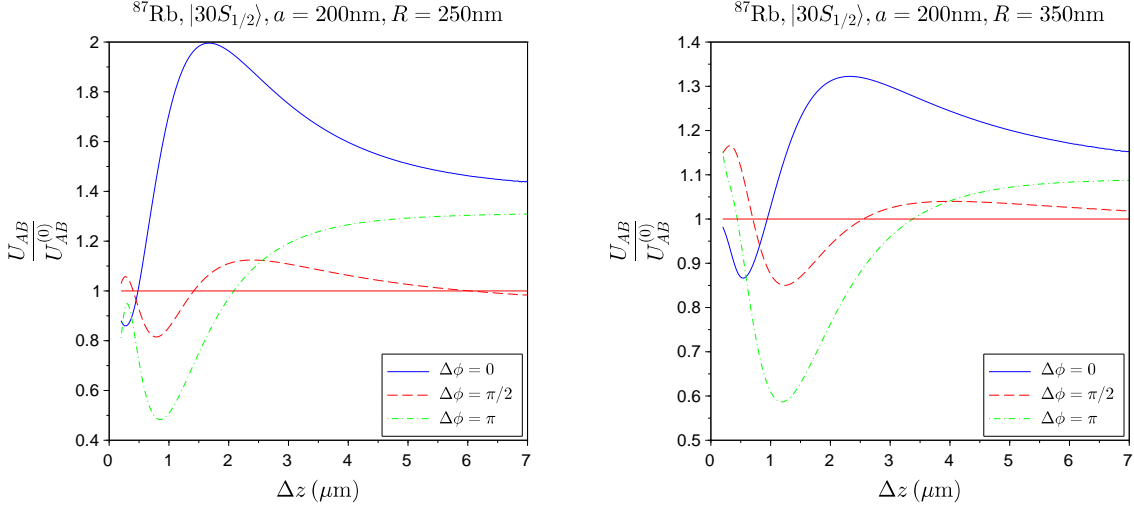}
\par\end{centering}
\caption{\textbf{Interaction between two $^{87}\mbox{Rb}$ atoms,
$\left(A,B\right)$, prepared in the state $\left|30S_{\nicefrac{1}{2}}\right\rangle $
in the neighbourhood of a nanofibre : influence of $\Delta\phi$.}
We consider $R_{A}=R_{B}=R$ and choose the quantisation axis along
$\left(Oz\right)$. We represent as functions of the lateral distance
$\Delta z$, the ratio $\left(\nicefrac{U_{AB}}{U_{AB}^{\left(0\right)}}\right)$
of van der Waals potentials between the two atoms located i)
at the distance $R=250\mbox{nm (left curve)},350\mbox{nm (right curve)}$
from the axis of a nanofibre of radius $a=200\mathrm{nm}$ $\left(\mbox{numerator }U_{AB}\right)$
and \emph{ii}) in free-space $\left(\mbox{denominator }U_{AB}^{\left(0\right)}\right)$
for different values of $\Delta\phi=0,\frac{\pi}{2},\pi$. The full red line corresponds to equality of both potentials.}

\label{FigRapportVideFibrePlusieursPhi}
\end{figure*}

In figure \ref{FigRapportVideFibrePlusieursPhi} we represented the
ratio $\left(\nicefrac{U_{AB}}{U_{AB}^{\left(0\right)}}\right)$ of
the potentials when atoms are \emph{i}) in the neighbourhood of a nanofibre $\left(\mbox{numerator }U_{AB}\right)$,
and \emph{ii}) in free-space $\left(\mbox{denominator }U_{AB}^{\left(0\right)}\right)$,
as a function of the lateral distance $\Delta z$. The fibre radius
is $a=200$nm, the two atoms are located at the same
distance from the fibre axis, i.e. $R=250$nm (left plot) and $R=350$nm
(right plot) and $\Delta\phi=0,\frac{\pi}{2},\pi$ .
At short distance, $\Delta z\lesssim10\times R$, the behaviour of
$\left(\nicefrac{U_{AB}}{U_{AB}^{\left(0\right)}}\right)$ strongly
varies from one configuration to another. The physical situation
is indeed very different, e.g., between $\Delta\phi=0$ and $\Delta\phi=\pi$
: \emph{i}) in the former case, when $\Delta z\rightarrow0$, $r_{AB}\rightarrow0$\footnote{The distance between the two atoms is given by $r_{AB}=\sqrt{\Delta z^{2}+4R^{2}\sin^{2}\frac{\Delta\phi}{2}}$.}
and the direct free-space interaction dominates, hence $\left(\nicefrac{U_{AB}}{U_{AB}^{\left(0\right)}}\right)\rightarrow1$
; \emph{ii}) in the latter case, even for $\Delta z=0$, $r_{AB}\neq0$
and the field reflected onto the fibre always plays an important role.
At large distance, i.e. for $\Delta z\gg R$, even though $r_{AB}\approx\Delta z$
and the quantisation and interatomic axes almost coincide for all
$\Delta\phi$'s, the contribution of the reflected field to the potential
strongly differs from one configuration to the other -- in general,
however, the presence of the nanofibre seems to enhance of the potential.
Until now, we were not able to design a simple model allowing us to
account for the features we observed : we can make the guess that
different values of $\Delta\phi$ favor the coupling of different
transitions to different modes of the reflected field. Numerical integration
issues, however, prevented us from pushing calculations to large values
of $\Delta z$ : even though the curves seem to tend towards an asymptote,
we neither were able to confirm this guess with certainty, nor could
we determine the hypothetical limiting value.

\section{Interaction between two atoms in the state $|nP_{3/2},M_{j}=\frac{3}{2}\rangle$\label{IntPP}}

As seen above, the presence of a nanofibre breaks the rotation symmetry of the tensor $\overline{{\bf T}}$ around the interatomic
axis, which is arbitrarily fixed along the quantisation axis. It also activates $\sigma^{\pm}-\sigma^{\pm}$-type couplings which are forbidden in
free-space. For $S$ states, studied in the previous section, these new
couplings do not modify the nature of the interaction between atoms.
It is quite different when atoms are prepared in a $P$ state as
we shall see below. To be more specific, in the following, we consider
that atoms are both prepared in the state $|nP_{3/2},M_{j}=\frac{3}{2}\rangle$.

We shall first investigate the dependence of the interaction potential
on the interatomic distance $\Delta z$ in the ``lateral'' configuration
$\Delta\phi=0$ and $R_{A}=R_{B}=R$, (Sec. \ref{PDeltaZ}). We will
show that the new couplings induced by the presence of the fibre\emph{
i}) strongly dominate the couplings allowed in free-space, \emph{ii})
strongly enhance the potential, and \emph{iii}) under certain conditions
can also make the potential attractive. We shall finally consider other
geometric configurations and study the dependence of the potentials with the
angle $\Delta\phi$ (Sec. \ref{PDeltaPhi}).

\subsection{Dependence on the lateral distance $\Delta z$\label{PDeltaZ}}
\begin{figure*}
\begin{centering}
\includegraphics[width=17cm]{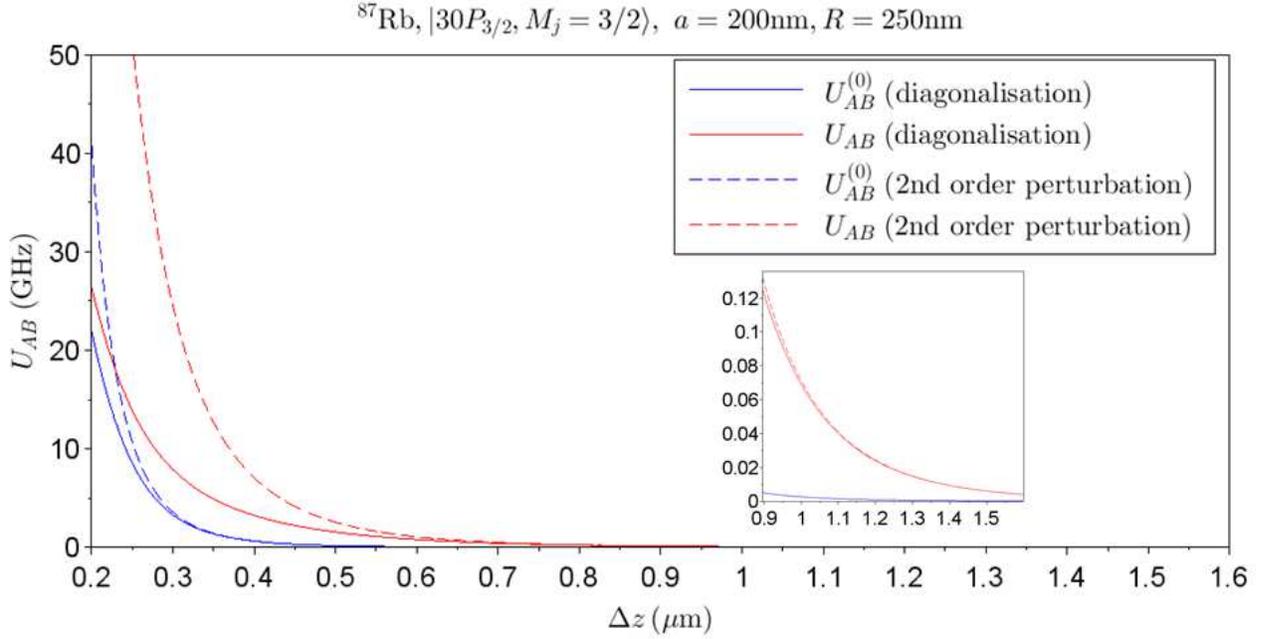}
\par\end{centering}
\caption{\textbf{Interaction between two $^{87}\mbox{Rb}$ atoms
$\left(A,B\right)$, prepared in the state $\left|30P_{\nicefrac{3}{2}},M_{j}=\frac{3}{2}\right\rangle $
in the neighbourhood of a nanofibre.} We fix $\Delta\phi=0$, $R_{A}=R_{B}=R$
and choose the quantisation axis along $\left(Oz\right)$. We plot
as functions of the lateral distance $\Delta z$ the van der
Waals interaction potentials between two atoms located i) at a distance
$R=250\mathrm{nm}$ of an optical nanofibre of radius $a=200\mathrm{nm}$,
$U_{AB}$ (red curves), and ii) in free-space, $U_{AB}^{\left(0\right)}$
(blue curves). Results presented here were obtained either \emph{i})
through direct diagonalisation of the Hamiltonian equation (\ref{HamiltonienComplet-1})
including the quadrupolar component (full-line curves), or \emph{ii})
via second-order perturbation theory relative to the same Hamiltonian
(dashed-line curves). The insert shows a zoom of the main plot on
the range $0.9\mu\mbox{m}\protect\leq\Delta z\protect\leq1.6\mu\mbox{m}$.}

\label{FigPotTot30P3}
\end{figure*}

In figure \ref{FigPotTot30P3}, we plotted as functions of $\Delta z$ the potentials i) in free-space, $U_{AB}^{\left(0\right)}$,
and \emph{ii}) at the distance $R=250$nm of a fibre of radius $a=200~\mathrm{nm}$,
$U_{AB}$. In free-space, the potential $U_{AB}^{\left(0\right)}$ is repulsive,
with the van der Waals radius $R_{\mathrm{vdW}}\approx0.25\mathrm{\mu m}$
and coefficient $C_{6}\left(|30P_{3/2},m_{j}=\frac{3}{2}\rangle\right)\approx-2.6\mathrm{MHz \cdot \left(\mu m\right)}^{6}$
for $\Delta z\gg R_{\mathrm{vdW}}$. Similarly to the states $|30S_{1/2}\rangle$,
the quadrupolar contribution is non negligible for $\Delta z<0.6\mathrm{\mu m}$,
though not dominant. By contrast, contrary to the case of $S$ states,
the potential is here always enhanced by the presence of the nanofibre,
and this increase is about one order of magnitude for $\Delta z>0.4~\mathrm{\mu m}$.
Moreover, here, the second-order perturbation theory is only valid
for $\Delta z\geq5R_{\mathrm{VdW}}$.

\begin{figure*}
\begin{centering}
\includegraphics[width=13cm]{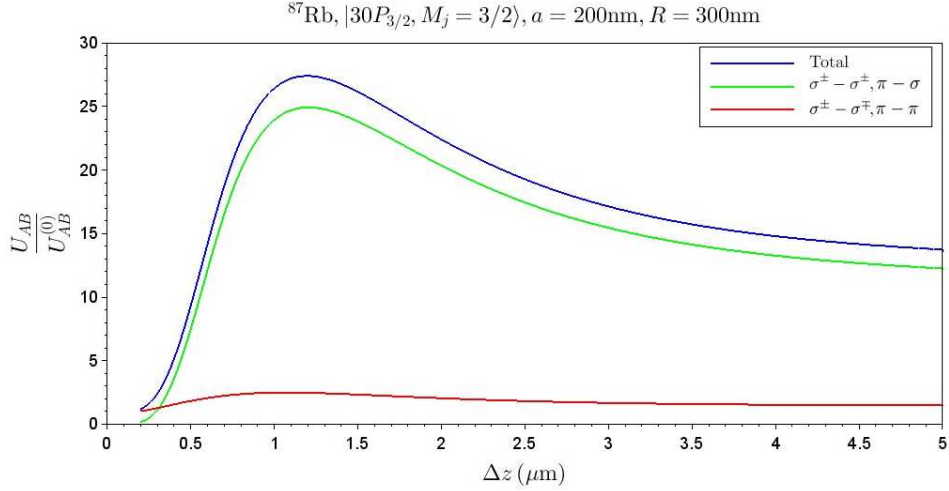}
\par\end{centering}
\caption{\textbf{Interaction between two  $^{87}\mbox{Rb}$ atoms,
$\left(A,B\right)$, prepared in the state $\left|30P_{\nicefrac{3}{2}},M_{j}=\frac{3}{2}\right\rangle $
in the neighbourhood of a nanofibre : partial contributions of the
couplings allowed and forbidden in the free-space.} We fix $\Delta\phi=0$,
$R_{A}=R_{B}=R$ and choose the quantisation axis along $\left(Oz\right)$.
We plotted $\left(\nicefrac{U_{AB}^{\left(\pi-\sigma\right)}+U_{AB}^{\left(\sigma^{\pm}-\sigma^{\pm}\right)}}{U_{AB}^{\left(0\right)}}\right)$
(green curve), $\left(\nicefrac{U_{AB}^{\left(\pi-\pi\right)}+U_{AB}^{\left(\sigma^{\pm}-\sigma^{\mp}\right)}}{U_{AB}^{\left(0\right)}}\right)$
(red curve), and $\frac{U_{AB}}{U_{AB}^{\left(0\right)}}$ (blue curve)
as functions of $\Delta z$ for $R=300$nm, where $U_{AB}^{\left(0\right)}$
is the van der Waals interaction potential between $A$ and
$B$ in free-space, $U_{AB}^{\left(\gamma\right)}$ is the partial contribution
of a coupling of type $\gamma=\pi-\pi,\pi-\sigma^{\pm},\sigma^{\pm}-\sigma^{\pm},\sigma^{\pm}-\sigma^{\mp}$
to the total van der Waals potential, $U_{AB}$, in the vicinity
of the fibre.}
\label{DetailContribution30P3}
\end{figure*}

To interpret this behaviour in the same spirit as in Sec. \ref{subsec:Apparition-de-nouveaux},
we plotted, in figure \ref{DetailContribution30P3}, the ratio $\left(\nicefrac{U_{AB}}{U_{AB}^{\left(0\right)}}\right)$
as a function of $\Delta z$ (blue curve) as well as the respective contributions
to this ratio of the couplings allowed in free-space, i.e. $\sigma^{\pm}-\sigma^{\mp}$
and $\pi-\pi$ (red curve), and of the new couplings induced by
the fibre, i.e. $\sigma^{\pm}-\sigma^{\pm}$ and $\pi-\sigma$ (green
curve). It appears that, contrary to the case of $S$ states, the
new couplings strongly dominate. More precisely, following the same
kind of analysis as in Sec. \ref{subsec:Modele-simplifie}
(cf figure \ref{FigDetailPertubation30S}), one identifies, in each
situation, the main coupling $\left|n_{A}\right\rangle \left|n_{B}\right\rangle \rightarrow\left|k_{A}\right\rangle \left|l_{B}\right\rangle $,
with\footnote{The condensed notation $|k\rangle_{A}\leftrightarrow|l\rangle_{B}$
designates the two states $\left\{ \left|a\right\rangle \left|b\right\rangle ,\left|b\right\rangle \left|a\right\rangle \right\} $
whose energy relative to the initial state is $\Delta_{kl}\equiv\omega_{nk}^{A}+\omega_{ml}^{B}$.}$\left|k_{A}\right\rangle \left|l_{B}\right\rangle =|30S_{1/2},M_{j}=\frac{1}{2}\rangle\leftrightarrow|29D_{5/2},M_{j}=\frac{5}{2}\rangle$
in free-space and $\left|k_{A}\right\rangle \left|l_{B}\right\rangle =|30S_{1/2},M_{j}=\frac{1}{2}\rangle\leftrightarrow|31S_{1/2},M_{j}=\frac{1}{2}\rangle$
in the neighbourhood of the fibre. The $\sigma^{+}-\sigma^{+}$-type
coupling, forbidden in free-space but allowed in the presence of the fibre,
strongly dominates due to the existence of a so-called (quasi) F\"orster
resonance.

\begin{figure*}
\begin{centering}
\includegraphics[width=8cm]{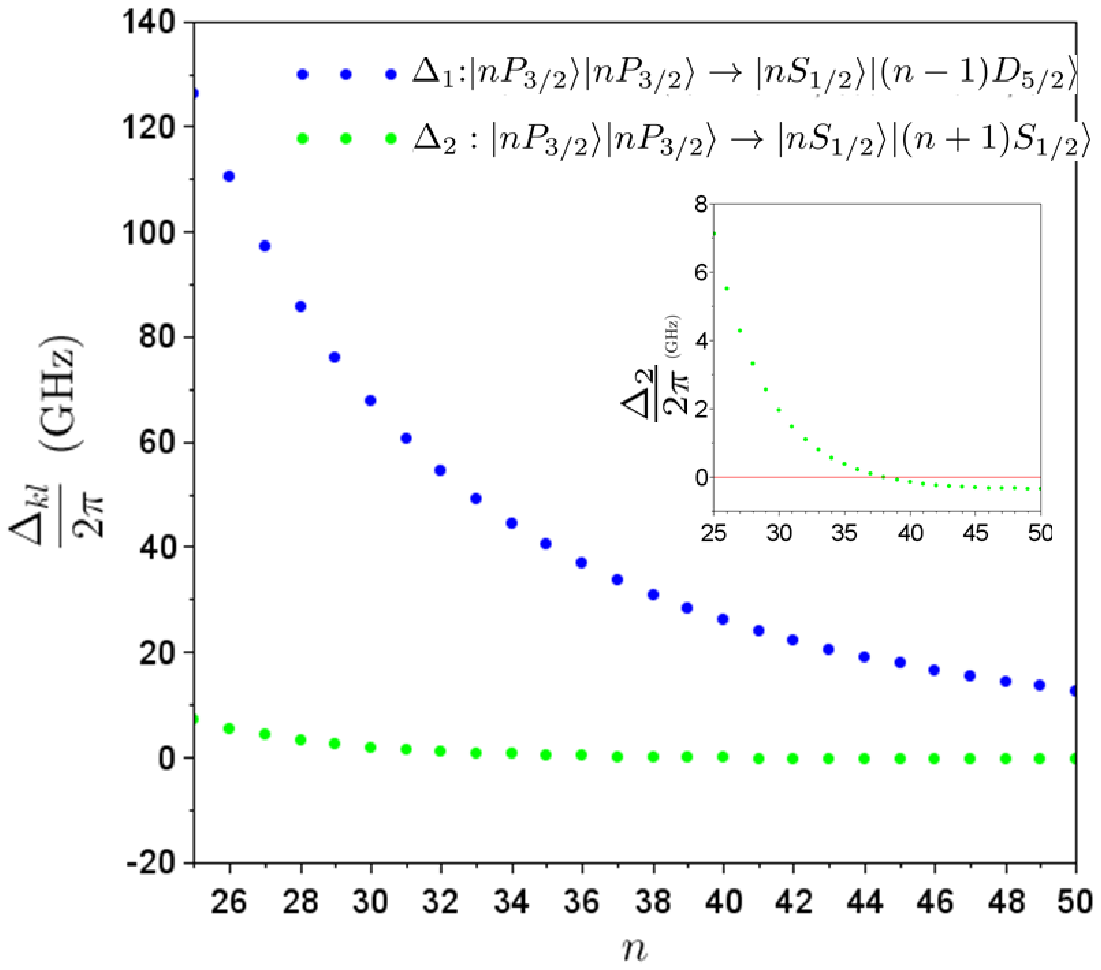}\includegraphics[width=8cm]{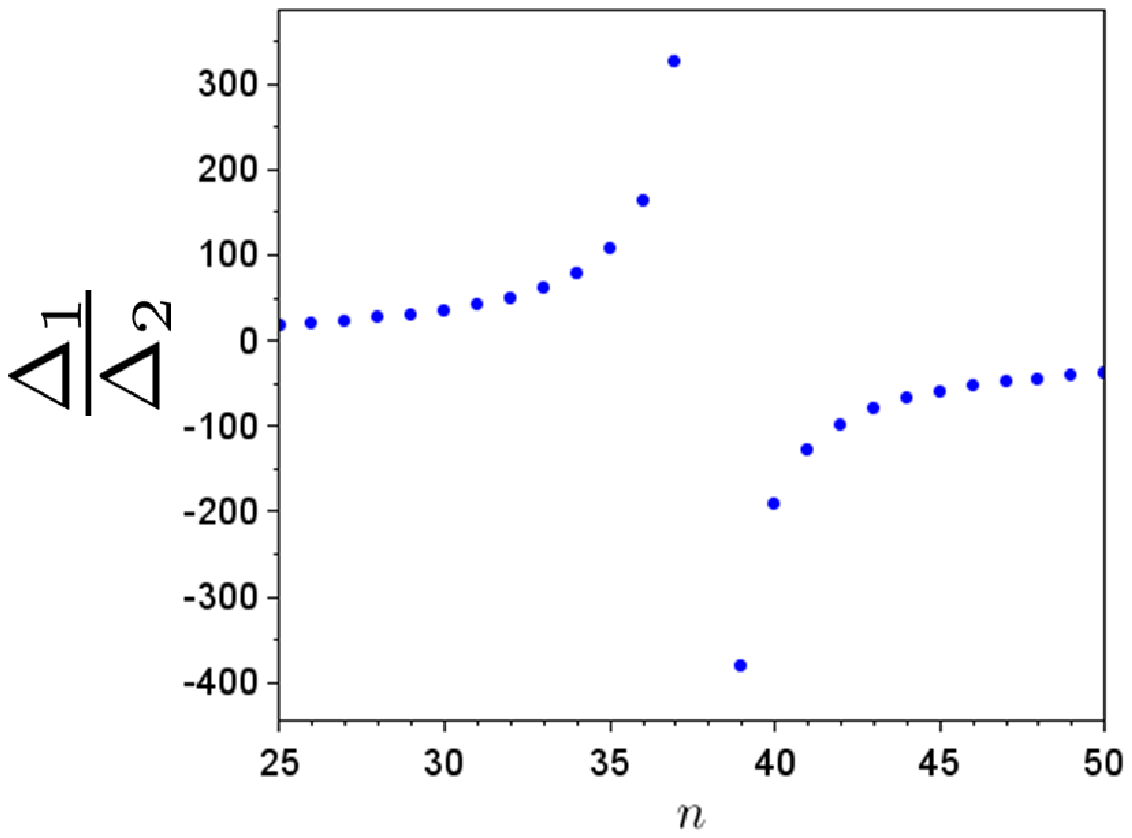}
\par\end{centering}
\caption{\textbf{Interaction between two  $^{87}\mbox{Rb}$ atoms,
$\left(A,B\right)$, prepared in the state $\left|nP_{\nicefrac{3}{2}},M_{j}=\frac{3}{2}\right\rangle $
in the neighbourhood of an optical nanofibre : existence of a (quasi)
}\textbf{F\"orster}\textbf{ resonance.} We fix $\Delta\phi=0$,
$R_{A}=R_{B}=R$ and choose the quantisation axis along $\left(Oz\right)$.
We plotted as functions of the principal quantum number, $n$, i)
the detunings $\Delta_{1}\left(n\right)$ and $\Delta_{2}\left(n\right)$
of the couplings $\left(1\right)$ and $\left(2\right)$ (cf main
text) which dominate the interaction potential in free-space and in the
neighbourhood of the fibre, respectively, (left panel), and ii) the
ratio $\nicefrac{\Delta_{1}}{\Delta_{2}}$ (right panel).}

\label{FigResForster}
\end{figure*}

To further investigate this point, we plotted, in figure \ref{FigResForster},
as functions of the principal quantum number, $n$, \emph{i}) the
detunings $\Delta_{1}\left(n\right)$ and $\Delta_{2}\left(n\right)$
of the transitions $\left(1\right)~\left|nP_{3/2}\right\rangle \left|nP_{3/2}\right\rangle  \rightarrow  \left|nS_{1/2}\right\rangle \left|\left(n-1\right)D_{5/2}\right\rangle$ and $\left(2\right)~\left|nP_{3/2}\right\rangle \left|nP_{3/2}\right\rangle  \rightarrow \left|nS_{1/2}\right\rangle \left|\left(n+1\right)S_{1/2}\right\rangle $
(figure \ref{FigResForster}, left panel), and \emph{ii}) the ratio $\nicefrac{\Delta_{1}\left(n\right)}{\Delta_{2}\left(n\right)}$
(figure \ref{FigResForster}, right panel). Coupling $\left(1\right)$
dominates the potential in free-space, $U_{AB}^{\left(0\right)}$,
while coupling $\left(2\right)$ dominates the potential
in the presence of the fibre, $U_{AB}$. We therefore have $U_{AB}^{\left(0\right)} \approx\frac{1}{\hbar\epsilon_{0}^{2}}\frac{\left|{\bf d}_{1}^{A}\cdot\overline{{\bf T}}_{0}\cdot{\bf d}_{1}^{B}\right|^{2}}{\Delta_{1}\left(n\right)}$ and $U_{AB} \approx\frac{1}{\hbar\epsilon_{0}^{2}}\frac{\left|{\bf d}_{2}^{A}\cdot\overline{{\bf T}}_{1}\cdot{\bf d}_{2}^{B}\right|^{2}}{\Delta_{2}\left(n\right)}$,
where ${\bf d}_{j=1,2}^{K=A,B}$ is the dipole operator of atom $K$
involved in the process $\left(j\right)$. We observe that, for $n=30$,
$\Delta_{2}\ll\Delta_{1}$ -- to be more explicit $\Delta_{2}\approx\frac{\Delta_{1}}{35}$. Assuming that $\left|{\bf d}_{1}^{A}\cdot\overline{{\bf T}}_{0}\cdot{\bf d}_{1}^{B}\right|^{2}$
and $\left|{\bf d}_{2}^{A}\cdot\overline{{\bf T}}_{1}\cdot{\bf d}_{2}^{B}\right|^{2}$
have the same order of magnitude, coupling $\left(2\right)$ therefore highly dominates coupling $\left(1\right)$
in the presence of the nanofibre, and the total potential is greatly
enhanced due to the presence of the fibre. Moreover, the ratio $\nicefrac{\Delta_{1}\left(n\right)}{\Delta_{2}\left(n\right)}$
-- and therefore the enhancement of the total potential in the presence
of the fibre -- first increases with $n$ up to $n=38$, at which
a so-called F\"orster (quasi-)resonance is observed, i.e. $\Delta_{2}\approx0$.
For $n>38$, $\Delta_{2}\left(n\right)$ becomes negative. The total
potential in the presence of the nanofibre, $U_{AB}\approx\frac{1}{\hbar\epsilon_{0}^{2}}\frac{\left|{\bf d}_{2}^{A}\cdot\overline{{\bf T}}_{1}\cdot{\bf d}_{2}^{B}\right|^{2}}{\Delta_{2}}$,
is \emph{repulsive} for $n<38$ and becomes \emph{attractive} for
$n>38$. By contrast, $\Delta_{1}\left(n\right)$ remains positive
and the potential in free-space, $U_{AB}^{\left(0\right)}\approx\frac{1}{\hbar\epsilon_{0}^{2}}\frac{\left|{\bf d}_{1}^{A}\cdot\overline{{\bf T}}_{0}\cdot{\bf d}_{1}^{B}\right|^{2}}{\Delta_{1}\left(n\right)}$,
remains repulsive on the considered range. For $n>38$, the presence
of the nanofibre hence modifies the nature of the van der Waals
force which suddenly becomes attractive.

\begin{figure*}
\begin{centering}
\includegraphics[width=16cm]{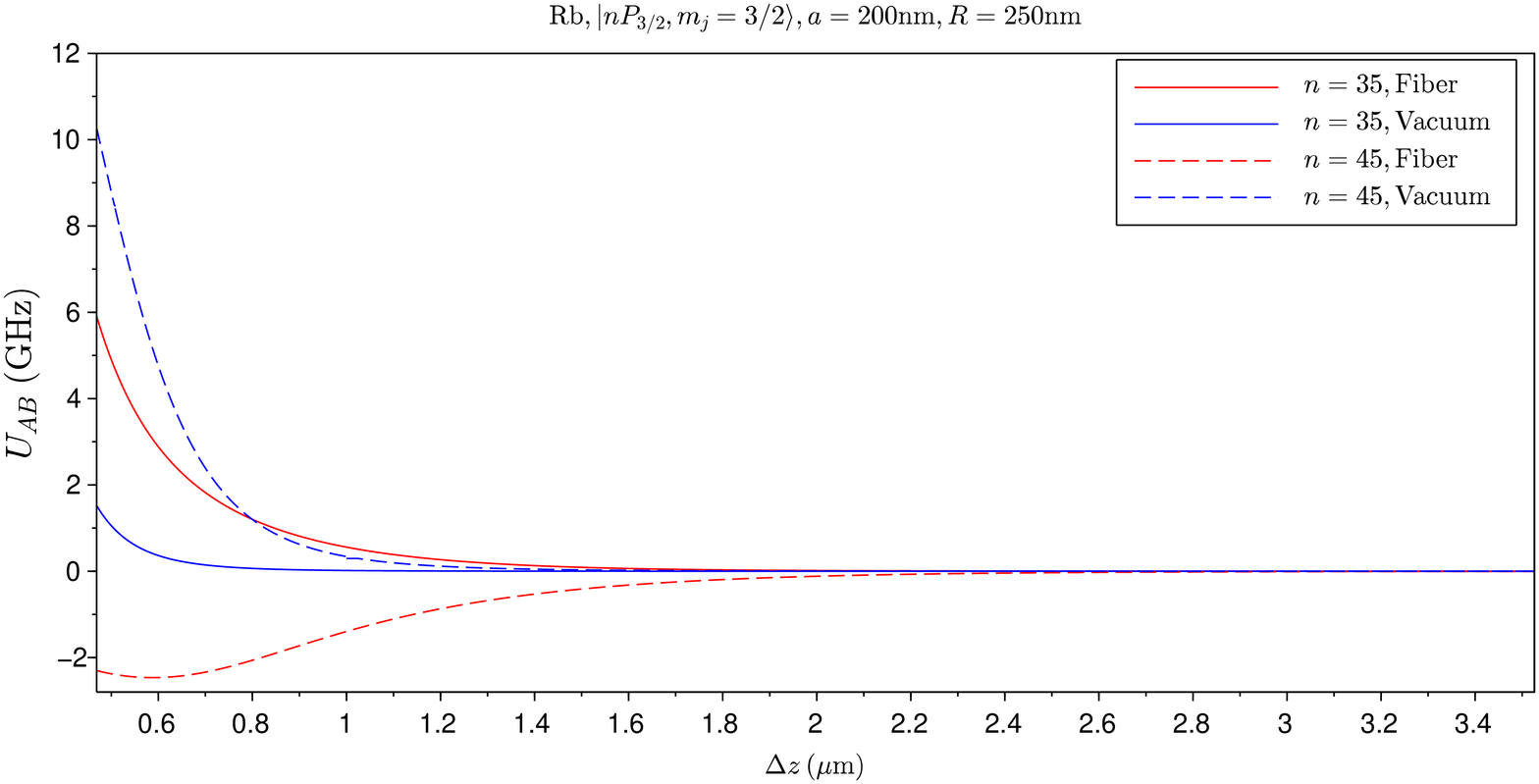}
\par\end{centering}
\caption{\textbf{Interaction between two $^{87}\mbox{Rb}$ atoms,
$\left(A,B\right)$, prepared in the state $\left|nP_{\nicefrac{3}{2}},M_{j}=\frac{3}{2}\right\rangle $
in the neighbourhood of a nanofibre : modification of the nature of
the potential close to a }\textbf{F\"orster}\textbf{ (quasi-)resonance.}
We fix $\Delta\phi=0$, $R_{A}=R_{B}=R$ and choose the quantisation
axis along $\left(Oz\right)$. We plotted as functions of the lateral
distance $\Delta z$ the van der Waals interaction potentials
between the atoms located i) at a distance $R=250\mathrm{nm}$ from
an optical nanofibre of radius $a=200\mathrm{nm}$, $U_{AB}$ (red
curves), and ii) in free-space, $U_{AB}^{\left(0\right)}$ (blue curves)
in the case $i$) $n=35$ (full-line curves) and \emph{ii}) $n=45$
(dashed-line curves). The results presented here were obtained through
direct diagonalisation of the Hamiltonian equation (\ref{HamiltonienComplet-1}).}
\label{FigPot35P45P}
\end{figure*}

This conclusion is confirmed and complemented by the results displayed
in figure \ref{FigPot35P45P}. The potential $U_{AB}$ is plotted as a function of $\Delta z$ when atoms are prepared in
the states $\left|35P_{3/2},M_{J}=\frac{3}{2}\right\rangle $ (full-line
curve) and $\left|45P_{3/2},M_{J}=\frac{3}{2}\right\rangle $ (dashed-line
curve), and located at the same distance $R_{A}=R_{B}=R=250\mathrm{nm}$
from the nanofibre. In free-space, both potentials are of repulsive nature, approximately scaling as $\nicefrac{C_{6}}{\Delta z^{6}}$
with $C_{6}\left(\left|35P_{3/2},M_{j}=\frac{3}{2}\right\rangle \right)\approx17~\mathrm{MHz.\left(\mathrm{\mu m}\right)^{6}}$
and $C_{6}\left(\left|45P_{3/2},M_{j}=\frac{3}{2}\right\rangle \right)\approx500~\mathrm{MHz.\left(\mathrm{\mu m}\right)^{6}}$
on the considered range of distances. For atoms prepared in the
state $\left|35P_{3/2},M_{J}=\frac{3}{2}\right\rangle $, $\Delta_{2}>0$
and the presence of the nanofibre enhances the potential $U_{AB}$
with respect to free-space (by a factor $\approx50$ for $\Delta z>1\mathrm{\mu m}$)
without changing its nature. By contrast, in the state $\left|45P_{3/2},M_{J}=\frac{3}{2}\right\rangle $,
$\Delta_{2}<0$, and the presence of the nanofibre therefore modifies
the nature of the potential which becomes attractive for $\Delta z>0.6\mu\mbox{m}$.
When atoms get closer, the effect of the fibre gets weaker and the
direct exchange of photons between atoms dominates : the total potential
hence becomes repulsive again and one observes the formation of a well whose minimum
is located around $\Delta z\approx0.6\mathrm{\mu m}$.

\subsection{Dependence on the angle $\Delta\phi$\label{PDeltaPhi}}

In the previous section, the observed enhancement of the interaction
between two atoms in the presence of an optical nanofibre was explained
by the appearance of a new and strongly dominating coupling, forbidden
in free-space but activated by the symmetry breaking induced by the fibre.
These results were obtained in the lateral configuration, i.e. for
$\Delta\phi=0$ et $R_{A}=R_{B}=R$ (I), and when the fibre, interatomic
and quantisation axes coincide with $\left(Oz\right)$ (II). Out of
this configuration, previous conclusions, \emph{a priori}, no longer hold. In particular,
the couplings which were forbidden in free-space under assumptions (I)
and (II) may become allowed and we therefore expect the relative enhancement
of the potential due to the introduction of the fibre less marked.

\begin{figure*}
\begin{centering}
\includegraphics[width=18cm]{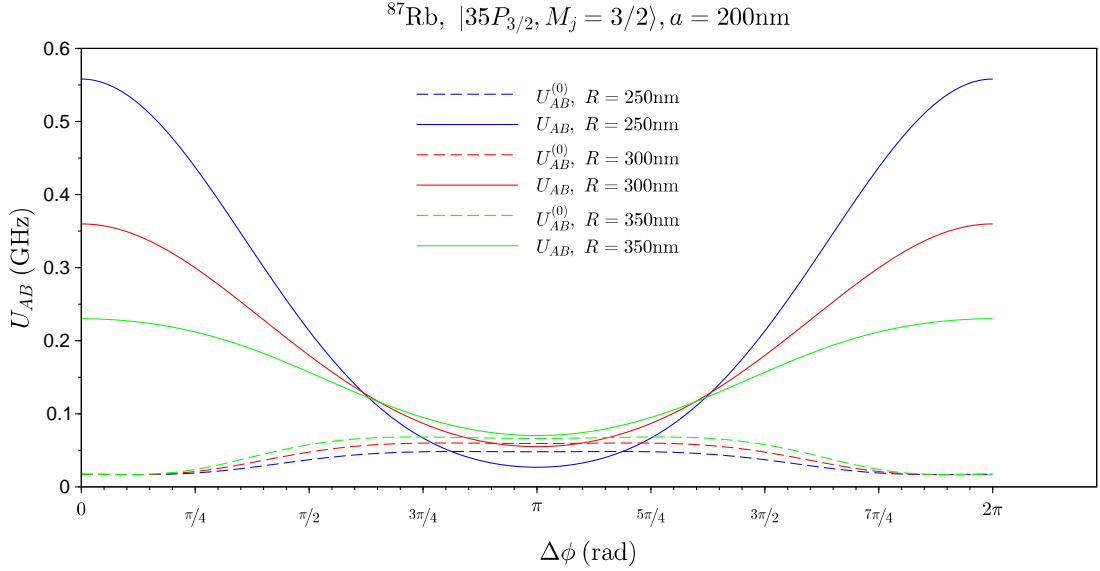}
\par\end{centering}
\caption{\textbf{Interaction between two $^{87}\mbox{Rb}$ atoms,
$\left(A,B\right)$, prepared in the state $\left|35P_{3/2},M_{j}=\frac{3}{2}\right\rangle $
in the neighbourhood of a nanofibre : influence of $\Delta\phi$.}
We fix $R_{A}=R_{B}=R$ and choose the quantisation axis along $\left(Oz\right)$.
We plotted as functions of $\Delta\phi$, the van der Waals
interaction potentials between the two atoms located i) in free-space,
$U_{AB}^{\left(0\right)}$ (dashed-line curves), and ii) in the neighbourhood
of a nanofibre of radius $a=200\mathrm{nm}$, $U_{AB}$ (full-line
curves). We fix $\Delta z=1\textrm{\ensuremath{\mu}m}$ and consider
three values for the distance of the atoms to the fibre axis, $R=250,300,350\mbox{nm}$.}

\label{FigDepPhi35P}
\end{figure*}

\begin{figure*}
\begin{centering}
\includegraphics[width=18cm]{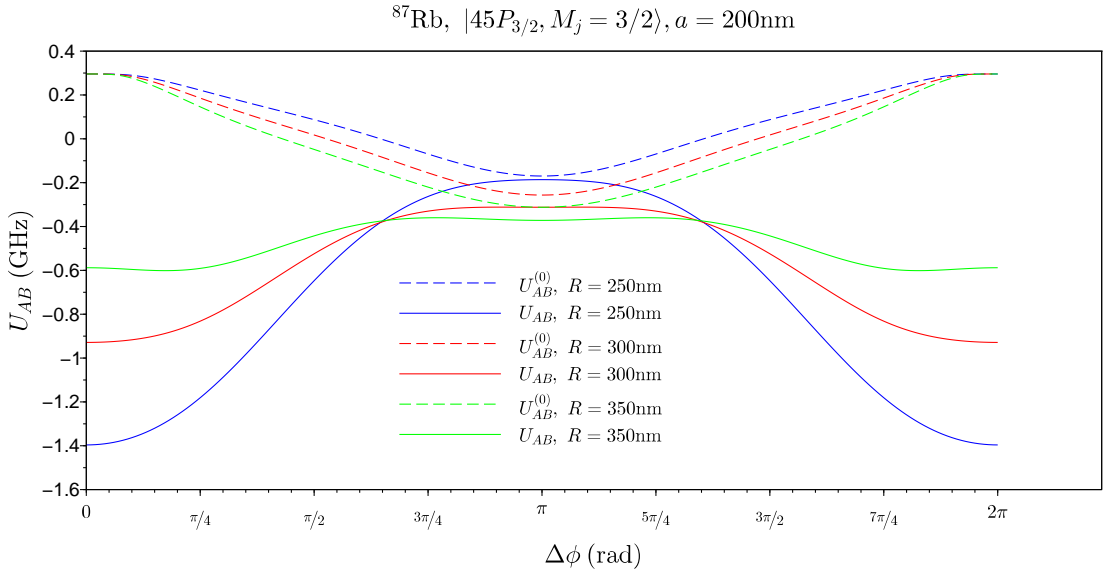}
\par\end{centering}
\caption{\textbf{Interaction between two $^{87}\mbox{Rb}$ atoms,
$\left(A,B\right)$, prepared in the state $\left|45P_{3/2},M_{j}=\frac{3}{2}\right\rangle $
in the neighbourhood of a nanofibre : influence of $\Delta\phi$.}
We fix $R_{A}=R_{B}=R$ and choose the quantisation axis along $\left(Oz\right)$.
We plotted as functions of $\Delta\phi$ the van der Waals
potentials between the atoms located i) in free-space, $U_{AB}^{\left(0\right)}$
(dashed-line curves), and ii) in the neighbourhood of a nanofibre
of radius $a=200\mathrm{nm}$, $U_{AB}$ (full-line curves). We fix
$\Delta z=1\textrm{\ensuremath{\mu}m}$ and consider three values
for the distance of the atoms to the fibre axis, $R=250,300,350\mbox{nm}$.}

\label{FigDepPhi45P}
\end{figure*}

To be more explicit, we plotted as functions of $\Delta\phi$ the potentials when atoms are located \emph{i}) in free-space, $U_{AB}^{\left(0\right)}$,
and\emph{ ii}) in the neighbourhood of a nanofibre, $U_{AB}$, separated
by a distance $\Delta z=1\mu\mbox{m}$ and both prepared in the states
$|35P_{3/2},M_{j}=\frac{3}{2}\rangle$ (figure \ref{FigDepPhi35P})
and $|45P_{3/2},M_{j}=\frac{3}{2}\rangle$ (figure \ref{FigDepPhi45P}),
for three values of the distance to the fibre axis $\left(Oz\right)$
$R_{A}=R_{B}=R=\left(250,300,350\right)\mbox{nm}$. The quantisation
axis is chosen along $\left(Oz\right)$ for all values of $\Delta\phi$. 

The system is symmetric with respect to the plane which contains atom
$A$ and $\left(Oz\right)$ axis, therefore $U_{AB}\left(\pi-\Delta\phi\right)=U_{AB}\left(\Delta\phi\right)$,
as can be seen in figures (\ref{FigDepPhi35P}, \ref{FigDepPhi45P}). The potential
is also obviously $2\pi$-periodic with $\Delta\phi$, i.e. $U_{AB}\left(\Delta\phi+2\pi\right)=U_{AB}\left(\Delta\phi\right)$.
We also underline that the potential in free-space, $U_{AB}^{\left(0\right)}$,
implicitly depends on $R$ through the interatomic distance $r_{AB}=\sqrt{\Delta z^{2}+4R^{2}\sin^{2}\frac{\Delta\phi}{2}}$,
and the inclination of the interatomic axis on the quantisation axis\footnote{Denoting by $\Theta$ the angle made by the interatomic and quantisation
axes, one has $\cos\Theta=\frac{\left(\frac{\Delta z}{R}\right)}{\sqrt{\left(\frac{\Delta z}{R}\right)^{2}+\sin^{2}\Delta\phi}}$.}.

For $n=35$, the potential $U_{AB}$ exhibits stronger variations
than $U_{AB}^{\left(0\right)}$ which always remains between $2.5$
and $6$ $\mathrm{GHz}$. For $0\leq\Delta\phi\leq\frac{\pi}{4}$
, the interaction potential is strongly enhanced by the presence of
the nanofibre, i.e. $\nicefrac{U_{AB}}{U_{AB}^{\left(0\right)}\gg1}$. This enhancement disappears in the range $\frac{\pi}{2}\leq\Delta\phi\leq\frac{3\pi}{2}$
where $U_{AB}$ becomes comparable with $U_{AB}^{\left(0\right)}$.
The same features are observed for $n=45$. The (negative) potential
$U_{AB}$ decreases in magnitude when $\Delta\phi$ increases from
0 to $\pi$. Around $\Delta \phi=\pi$, $U_{AB}$ and $U_{AB}^{\left(0\right)}$
have the same order of magnitude and sign. In particular, these plots
show that the sign change of the potential induced by the presence
of the nanofibre, previously observed for $\Delta\phi=0$, actually
extends to the range $0\leq\Delta\phi\leq\frac{\pi}{2}$.

\section{Rotation of the quantisation axis\label{Anisotropy}}

The van der Waals interaction between two atoms in free-space is, \emph{a priori}, anisotropic. According to equation (\ref{eq:ExpressionC6-1}), the potential $U_{AB}^{\left(0\right)}$
indeed depends on the relative direction of the interatomic and quantisation
axes. To be more explicit, denoting by ${\bf u}_{AB}\equiv\frac{{\bf r}_{B}-{\bf r}_{A}}{\left|{\bf r}_{B}-{\bf r}_{A}\right|}$ and ${\bf e}_{\mbox{q}}$
the respective unit vectors of these axes and $\Theta$ the angle they form $\left({\bf e}_{\mbox{q}}\cdot{\bf u}_{AB}=\cos\Theta\right)$, one has $C_{6}^{\left(0\right)}=C_{6}^{\left(0\right)}\left(\Theta\right)$. This anisotropy was demonstrated experimentally and its influence on the
Rydberg blockade investigated \cite{BRL14}.

\begin{figure*}
\begin{centering}
\includegraphics[width=10cm]{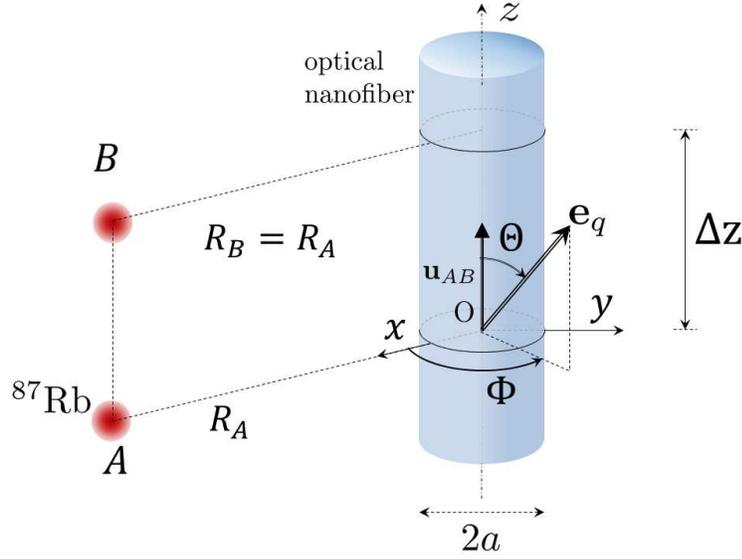}
\par\end{centering}
\caption{\textbf{Two $^{87}\mbox{Rb}$ atoms, $\left(A,B\right)$,
in the neighbourhood of a silica optical nanofibre : quantisation axis
of arbitrary direction.} We use the same Cartesian frame as in figure
\ref{SchemaDeuxAtomes} and we fix $\Delta\phi=0$ and $R_{A}=R_{B}=R$,
so that the interatomic axis is parallel to the fibre axis taken as
$\left(Oz\right)$ axis. The quantisation axis has the unit vector
${\bf e}_{\mbox{q}}$ defined by its spherical coordinates $\left(\Theta,\Phi\right)$,
i.e. ${\bf e}_{\mbox{q}}=\sin\Theta\cos\Phi\;{\bf e}_{x}+\sin\Theta\sin\Phi\;{\bf e}_{y}+\cos\Theta\;{\bf e}_{z}$.}
\label{FigAngQuant2Atomes}
\end{figure*}

The presence of an optical nanofibre brings a new priviledged direction,
i.e. the fibre axis, which is conventionally taken as $\left(Oz\right)$ axis.
Until now, we fixed the quantisation axis used to define atomic states along
$\left(Oz\right)$, i.e. we assumed the atomic dipoles pointed along the same direction $Oz$, and studied how changing the direction of the interatomic axis modifies the interaction potential (Secs. \ref{DependanceDeltaPhi},
\ref{PDeltaPhi}). By contrast, in this section, we shall assume the interatomic
axis along $\left(Oz\right)$, fix $\Delta\phi=0$ and $R_{A}=R_{B}=R$,
and shall consider that quantisation axis is along an arbitrary unit vector ${\bf e}_{\mbox{q}}$ defined by the angles $\left(\Theta,\Phi\right)$
(see figure \ref{FigAngQuant2Atomes}). Note that the rotation symmetry 
around the interatomic axis which exists in free-space is no longer checked
near  the nanofibre. The $C_{6}$ coefficient,
\emph{a priori}, depends not only on the angle $\Theta$ but
also on $\Phi$.

We first study how the rotation of the quantisation axis
modifies the interaction potential when atoms are prepared in the state
$\left|30P_{3/2},M_{j}=\frac{3}{2}\right\rangle $ (Sec. \ref{30P3/23/2}).
We qualitatively reproduce the results obtained \emph{via} a simplified
model, restricted to a single $\sigma^{+}-\sigma^{+}$-type coupling,
which allows us to relate the modification of the potential to the fibre-induced
symmetry breaking (Sec. \ref{ModelSigSig}).

\subsection{State $\left|30P_{3/2},M_{j}=\frac{3}{2}\right\rangle $\label{30P3/23/2}}

\begin{figure*}
\begin{centering}
\includegraphics[width=17cm]{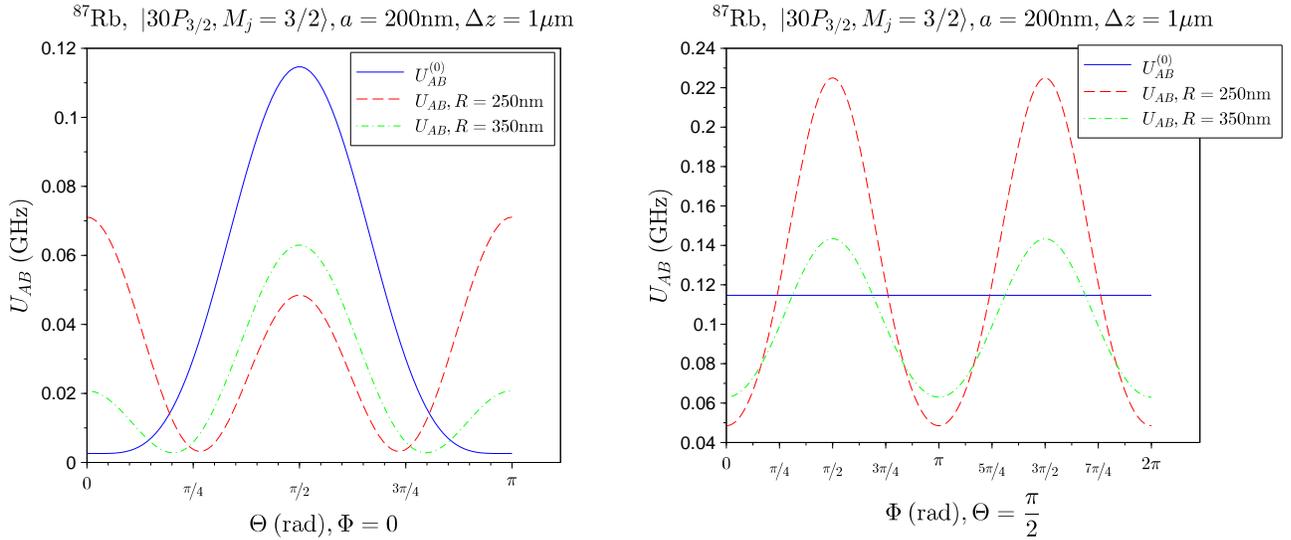}
\par\end{centering}
\caption{\textbf{Interaction between two  $^{87}\mbox{Rb}$ atoms,
$\left(A,B\right)$, prepared in the state $\left|30P_{3/2},M_{j}=\frac{3}{2}\right\rangle $
in the neighbourhood of a nanofibre : influence of the rotation of
the quantisation axis.} We fix $\Delta\phi=0$ and $R_{A}=R_{B}=R$.
We plotted the van der Waals potentials between the atoms i)
in free-space, $U_{AB}^{\left(0\right)}$ (full-line curves), and ii)
in the neighbourhood of a nanofibre of radius $a=200\mathrm{nm}$,
$U_{AB}$ (dashed-line curves), as functions of \emph{i}) $\Theta$
for $\Phi=0$, i.e. for the quantisation axis rotating in the plane
$\left(Oxz\right)$ (left panel), and\emph{ ii}) $\Phi$ for $\Theta=\frac{\pi}{2}$,
i.e. for the quantisation axis rotating in the plane $\left(Oxy\right)$
(right panel). We fix $\Delta z=1\textrm{\ensuremath{\mu}m}$ and
consider two values for the distance of the atoms from the $\left(Oz\right)$
axis, $R=250,350\mbox{nm}$.}
\label{DepQuant30P3}
\end{figure*}

In figure \ref{DepQuant30P3} we plotted the potentials
in free-space, $U_{AB}^{\left(0\right)}$, and in the neighbourhood of
a nanofibre, $U_{AB}$, when atoms are prepared in the state $|30P_{3/2},M_{j}=\frac{3}{2}\rangle$,
as functions of $\Theta$ for $\Phi=0$ (left panel) and of $\Phi$
for $\Theta=\frac{\pi}{2}$ (right panel). The radius of the nanofibre
is $a=200~\mathrm{nm}$, the interatomic lateral distance is $\Delta z=1\mathrm{\mu m}$,
$\Delta\phi=0$ and we consider two values for the distance of atoms
to the $\left(Oz\right)$ axis, $R_{A}=R_{B}=R=250,350\mbox{nm}$. 

As a function of $\Theta$, the free-space potential  $U_{AB}^{\left(0\right)}\left(|30P_{3/2},M_{j}=\frac{3}{2}\rangle\right)$,
is maximal (resp. minimal) in $\Theta=\frac{\pi}{2}$ (resp. $\Theta=\left(0,\pi\right)$), i.e. when dipoles point orthogonally to (resp. are along) the interatomic axis. As expected, $U_{AB}^{\left(0\right)}\left(|30P_{3/2},M_{j}=\frac{3}{2}\rangle\right)$
does not depend on $\Phi$.

In the presence of the nanofibre, the potential, $U_{AB}$, behaves in a quite different manner. 

A. When the quantisation axis rotates in the plane $\left(Oxz\right)$,
i.e. when $\Theta$ varies and $\Phi=0$ (left panel) : \emph{i})
if $R$ is weak enough ($250\mathrm{nm}$), the absolute minimum reached in free-space for ${\bf e}_{q}={\bf e}_{z}$ becomes the absolute maximum
; if $R$ increases, this maximum remains, though local. \emph{ii})
The potential maximum in free-space, reached in $\Theta=\frac{\pi}{2}$,
i.e. when ${\bf e}_{q}={\bf e}_{x}$, remains a local maximum in the presence of the nanofibre, though much weaker ; \emph{iii}) potential
minima in the presence of the fibre are reached in $\Theta=\frac{\pi}{4},\frac{3\pi}{4}$,
by contrast with free-space where these minima are achieved
in $\Theta=0,\pi$.

B. When the quantisation axis rotates in the plane $\left(Oxy\right)$
around the interatomic axis $\left(Oz\right)$, i.e. when $\Phi$
varies and $\Theta=\frac{\pi}{2}$ (right panel), the potential minimum
is reached in $\Phi=0$, i.e. when ${\bf e}_{q}={\bf e}_{x}$, while the
(absolute) maximum is reached in $\Phi=\frac{\pi}{2}$,
i.e. when ${\bf e}_{q}={\bf e}_{y}$. This second plot shows that
the nanofibre-induced symmetry breaking causes the potential to depend
on the angle $\Phi$.

To qualitatively account for these results, we develop below a simplified model restricted to a single coupling, in the same spirit as in Sec.
\ref{subsec:Modele-simplifie}.

\subsection{Simplified model restricted to a $\sigma^{+}-\sigma^{+}$ coupling\label{ModelSigSig}}

The potential $U_{AB}$ is found to be highly dominated by the following
$\sigma^{+}-\sigma^{+}$-type coupling, 
$\left|30P_{3/2},M_{j}=\frac{3}{2}\right\rangle \left|30P_{3/2},M_{j}=\frac{3}{2}\right\rangle 
\rightarrow \left|30S_{1/2},M_{j}=\frac{1}{2}\right\rangle \left|31S_{1/2},M_{j}=\frac{1}{2}\right\rangle$,
already identified in the previous section. It is also true for the potential in free-space, $U_{AB}^{\left(0\right)}$, except around ${\bf e}_{q}={\bf e}_{z}$. In the rest of this section, we shall restrict ourselves to this single
coupling.

We assume the quantisation axis is defined by the angles $\left(\Theta,\Phi\right)$
(figure \ref{FigAngQuant2Atomes}), and the dipoles ${\bf d}_{A}^{\sigma^{+}}$
and ${\bf d}_{B}^{\sigma^{+}}$ take the following form
\[
\left(\begin{array}{c}
d_{x}\\
d_{y}\\
d_{z}
\end{array}\right)_{\sigma^{+}}=\frac{d^{\sigma^{+}}}{\sqrt{2}}\left(\begin{array}{c}
-\cos\Theta\cos\Phi-\mathrm{i}\sin\Phi\\
-\cos\Theta\sin\Phi+\mathrm{i}\cos\Phi\\
\sin\Theta
\end{array}\right)
\]
The non-retarded Green's functions, ${\bf \overline{T}}_{0}\left({\bf r}_{A},{\bf r}_{B}\right)$
and ${\bf \overline{T}}_{1}\left({\bf r}_{A},{\bf r}_{B}\right)$,
Eqs. (\ref{eq:GreenVideConfigLateral},\ref{eq:GreenFibConfigLateral}),
lead to
\begin{eqnarray*}
{\bf d}_{A}^{\sigma^{+}}\cdot{\bf \overline{T}}_{0}\cdot{\bf d}_{B}^{\sigma^{+}} & =\frac{d_{A}^{\sigma^{+}}d_{B}^{\sigma^{+}}}{2}\frac{3}{4\pi}\frac{1}{\left(\Delta z\right)^{3}}\sin^{2}\Theta\\
{\bf d}_{A}^{\sigma^{+}}\cdot{\bf \overline{T}}_{1}\cdot{\bf d}_{B}^{\sigma^{+}} & =\frac{d_{A}^{\sigma^{+}}d_{B}^{\sigma^{+}}}{2}\left[T_{m}\sin^{2}\Theta\right.\\
 & \left.+\Delta T\left(1+\cos^{2}\Theta\right)\cos2\Phi-\mathrm{i}\Delta T\cos\Theta\sin2\Phi\right]
\end{eqnarray*}
where we set $\Delta T\equiv\frac{1}{2}\left(\left[{\bf \overline{T}}_{1}\right]_{xx}-\left[{\bf \overline{T}}_{1}\right]_{yy}\right)$,
$T_{m}\equiv\left[{\bf \overline{T}}_{1}\right]_{zz}-\frac{1}{2}\left(\left[{\bf \overline{T}}_{1}\right]_{xx}+\left[{\bf \overline{T}}_{1}\right]_{yy}\right)$
and ${\bf \overline{T}}_{1}\equiv\sum_{ij=x,y,z}\left[{\bf \overline{T}}_{1}\right]_{ij}{\bf e}_{i}\otimes{\bf e}_{j}$.
Note that \emph{i}) $\Delta T$ characterizes the nanofibre-induced symmetry breaking and \emph{ii}) $\Delta T$ and $T_{m}$ do not depend on the quantisation axis direction but only on the fibre geometric characteristics and the positions of the atoms.

We first deduce that $U_{AB}^{\left(0\right)}\propto\left|{\bf d}_{A}^{\sigma^{+}}\cdot{\bf \overline{T}}_{0}\cdot{\bf d}_{B}^{\sigma^{+}}\right|^{2}\propto\sin^{4}\Theta$
which agrees with the results displayed in figure \ref{DepQuant30P3}.
In particular, in free-space, the potential does not depend on $\Phi$, vanishes when
$\Theta=0$, and reaches its maximum in $\Theta=\frac{\pi}{2}$. 

Moreover, we get
\begin{eqnarray}
\left|{\bf d}_{A}^{\sigma^{+}}\cdot\left({\bf \overline{T}}_{1}+\overline{{\bf T}}_{0}\right)\cdot{\bf d}_{B}^{\sigma^{+}}\right|^{2} & =\frac{\left(d_{A}^{\sigma^{+}}d_{B}^{\sigma^{+}}\right)^{2}}{4}\label{dTd}\\
 & \times\left\{ \begin{array}{c}
\left(T_{0}+T_{m}-\Delta T\right)^{2}\left(\sin^{2}\Theta-\eta_{1}\right)^{2}\quad\mbox{for }\Phi=0\\
\left(T_{0}+T_{m}\right)^{2}\left(1-\eta_{2}\cos2\Phi\right)^{2}\quad\mbox{for }\Theta=\frac{\pi}{2}
\end{array}\right.\nonumber 
\end{eqnarray}
where we set $T_{0}\equiv\frac{3}{4\pi}\frac{1}{\left(\Delta z\right)^{3}}$,
$\eta_{1}\equiv-\frac{2\Delta T}{T_{0}+T_{m}-\Delta T}$ and $\eta_{2}\equiv-\frac{\Delta T}{T_{0}+T_{m}}$.
We shall see that this formula enables us to account for the features
of the potential in the neighbourhood of the nanofibre, $U_{AB}\propto\left|{\bf d}_{A}^{\sigma^{+}}\cdot\left({\bf \overline{T}}_{1}+\overline{{\bf T}}_{0}\right)\cdot{\bf d}_{B}^{\sigma^{+}}\right|^{2}$.
We underline that in the lateral configuration, i.e. for $\Delta\phi=0$
and $R_{A}=R_{B}=R$, $\Delta T$ is always negative for $\Delta z$
large enough, i.e. $\Delta z>270\mathrm{nm}$ (resp. $500\mathrm{nm}$)
for $R=250\mathrm{nm}$ (resp. $350\mathrm{nm}$).

\subsection*{Dependence on $\Theta$ for $\Phi=0$}

For $\Phi=0$, the potential $U_{AB}$, as a function of $\Theta$,
varies as $\left(\sin^{2}\Theta-\eta_{1}\right)^{2}$ which has two
local maxima in $\Theta=0,\frac{\pi}{2}$, respectively $A_{1}=\eta_{1}^{2}$
and $A_{2}=\left(1-\eta_{1}\right)^{2}$, and two zeroes in $\Theta_{\mathrm{min}}=\arcsin\left(\sqrt{\eta_{1}}\right)$
et $\Theta=\pi-\Theta_{\mathrm{min}}$.

\begin{table}
\begin{centering}
\begin{tabular}{|c|c|c|c|}
\hline 
 & $A_{1}$ & $A_{2}$ & $\Theta_{\min}$ (°)\tabularnewline
\hline 
$R=250\mathrm{nm}$ & 0.52 & 0.08 & 63\tabularnewline
\hline 
$R=350\mathrm{nm}$ & 0.16 & 0.35 & 39\tabularnewline
\hline 
\end{tabular}
\par\end{centering}
\label{TabA1A2Thetamin}

\caption{\textbf{Function $\left(\sin^{2}\Theta-\eta_{1}\right)^{2}$}~: maxima
$A_{1}$ and $A_{2}$ reached in $\Theta=0$ and $\Theta=\frac{\pi}{2}$,
zero $\Theta_{\mathrm{min}}$. These values are calculated from the
Green's function of the fibre for $\Delta z=1\mu\mathrm{m}$
and $R_{A}=R_{B}=R=250,350\mathrm{nm}$ .}
\end{table}

The values $\left(A_{1},A_{2},\Theta_{\mathrm{min}}\right)$ given
in Table \ref{TabA1A2Thetamin} were obtained for a system of two
atoms separated by $\Delta z=1\mathrm{\mu m}$ and located
at the distance $R_{A}=R_{B}=R=\left(250,350\right)\mathrm{nm}$ from
the fibre axis. These values allow us to account for the behaviour
of the potential $U_{AB}$ represented in figure \ref{DepQuant30P3}
(left panel). For $R=350\mathrm{nm}$, $A_{1}<A_{2}$, and, as expected
from equation (\ref{dTd}), the local maximum in $\Theta=0$ is less marked
than the maximum in $\Theta=\frac{\pi}{2}$. By contrast, for $R=250\mathrm{nm}$,
$A_{1}>A_{2}$, and the opposite behaviour is observed. We also recover
the position of the minimum~: for $R=250\mathrm{nm}$ (resp. $350\mbox{nm})$,
it is reached in $\Theta_{\mathrm{min}}>\frac{\pi}{4}$ (resp. $<\frac{\pi}{4}$).

We underline that the departure between the potentials in the neighbourhood
of the nanofibre and in free-space is governed by the term $\eta_{1}$.
For $\eta_{1}\rightarrow0$, one has $A_{1}\rightarrow0$, $A_{2}\rightarrow1$
and $\Theta_{\mathrm{min}}\rightarrow0$. For $\eta_{1}>0.5$, the
maximum in $\Theta=0$ becomes the absolute maximum and the maximum
in $\frac{\pi}{2}$ becomes local. Finally, for $\eta_{1}\rightarrow1$,
the profile is inversed, $A_{1}\rightarrow4$, $A_{2}\rightarrow0$,
and the potential zero is achieved in $\Theta_{\mathrm{min}}=\frac{\pi}{2}$. 

\subsection*{Dependence on $\Phi$ for $\Theta=\frac{\pi}{2}$}

For $\Theta=\frac{\pi}{2}$, the potential $U_{AB}$, as a function
of $\Phi$, varies as $\left(1+\eta_{2}\cos2\Phi\right)^{2}$. For
$\Delta T<0$, this function has a minimum in $\Phi=0$ and a maximum
in $\Phi=\frac{\pi}{2}$, which indeed corresponds to the behaviour
observed for $U_{AB}$ in figure \ref{DepQuant30P3} (right panel).
We underline that the departure between the potentials in the neighbourhood
of the nanofibre and in free-space -- in the latter case, $U_{AB}^{\left(0\right)}$
does not depend on $\Phi$ -- is governed by the term $\eta_{2}$.

\subsection*{General case}

\begin{figure*}
\begin{centering}
\includegraphics[width=19cm]{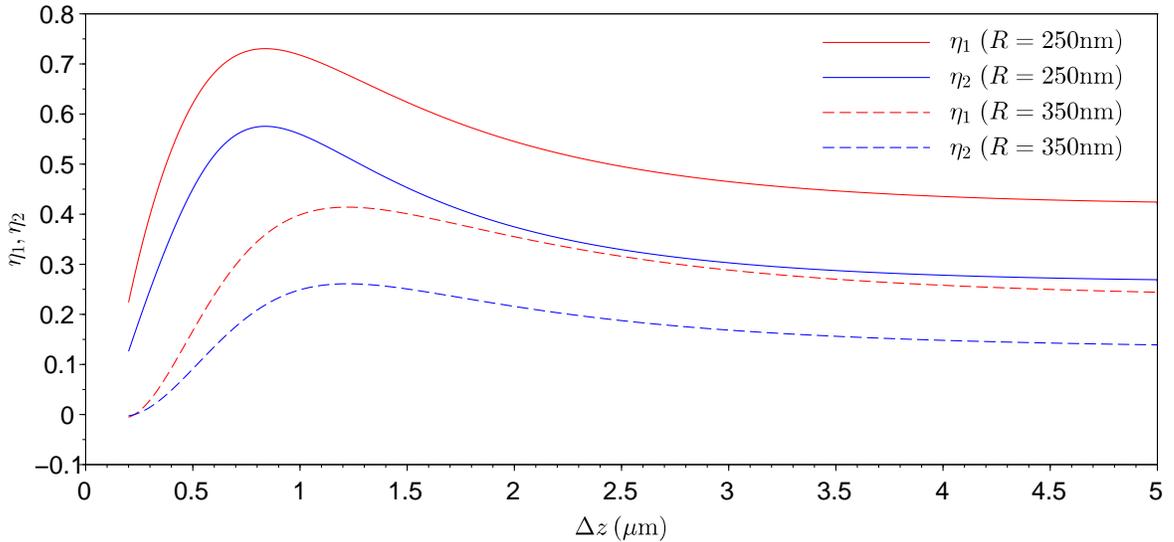}
\par\end{centering}
\caption{\textbf{Coefficients $\left(\eta_{1},\eta_{2}\right)$ which characterize
the discrepancy between the interaction potentials in free-space and in the neighbourhood
of a nanofibre : dependence on the interatomic lateral
distance, $\Delta z$, and the distance of the atoms to the fibre
axis, $R$.} We fix $\Delta\phi=0$ and $R_{A}=R_{B}=R$. We plotted
the coefficients $\eta_{1}=-\frac{2\Delta T}{T_{0}+T_{m}-\Delta T}$
(red curves) and $\eta_{2}=-\frac{\Delta T}{T_{0}+T_{m}}$ (blue curves)
as functions of the interatomic lateral distance $\Delta z$ for two
values of the distance $R=350\mathrm{nm}$ (full-line curves) and
$250\mathrm{nm}$ (dashed-line curves).}

\label{DepEtaEnZ}
\end{figure*}

As seen above, coefficients $\left(\eta_{1},\eta_{2}\right)$ characterize
the discrepancy between the interaction potentials in free-space and in the neighbourhood
of the nanofibre. For weak $\eta_{1}$, potential $U_{AB}$ as a function
of $\Theta$ approximately varies as $\sin^{4}\Theta$. For weak $\eta_{2}$,
$U_{AB}$ does not depend on $\Phi$. 

In the general case, i.e. for arbitrary $\Theta$ and $\Phi$, these
two coefficients simultaneously come into play.

To complement our discussion, we plotted in figure \ref{DepEtaEnZ}
coefficients $\eta_{1}$ and $\eta_{2}$ as functions of 
$\Delta z$ for $R=250\mathrm{nm}$ (full-line curves) and $R=350\mathrm{nm}$
(dashed-line curves). For each value of $R$, there exists a certain
distance $\Delta z_{\mbox{max}}$ around which the discrepancy between $U_{AB}^{\left(0\right)}$ and $U_{AB}$
is the most marked. From the plot, one gets $\Delta z_{\mbox{max}}\approx0.7\mathrm{\mu m}$
$\left(\mbox{resp. }1.2\mathrm{\mu m}\right)$ for $R=250\mathrm{nm}$
$\left(\mbox{resp. }350\mathrm{nm}\right)$. When atoms are too close, i.e. $\Delta z<\Delta z_{\mbox{max}}$, this discrepancy gets
weaker. In the same way, when $\Delta z\rightarrow+\infty$, $\eta_{1}$
and $\eta_{2}$ slowly decrease, seemingly towards a limiting value
-- though we were not yet able to prove it -- which is higher for
lower values of $R$.

\section{Conclusion\label{Conclusion}}

In this article, we theoretically investigated the van der Waals interaction of two Rydberg rubidium atoms $^{87}$Rb in the presence of a silica optical nanofibre. In the case of S states, when interatomic and fibre axes are parallel, the repulsive potential is enhanced (resp. decreased) at long (resp. short) interatomic distances with respect to free-space, and blockade radius is enhanced. The ratio between the potentials in free-space and in the presence of the nanofibre moreover does not depend on $n$ at large distance. Restricting ourselves to dominating couplings we could account for the main features observed and relate them to the activation of new couplings -- forbidden in free-space -- due to the fibre-induced breaking of the rotation symmetry around the interatomic axis.
In the case of P Rydberg states, we showed the interaction potential is always increased by the presence of the nanofibre. New couplings induced by the nanofibre-assisted-symmetry-breaking now dominate due to the existence of a Förster quasi-resonance. They may even make the potential attractive on some distance range, therefore leading to the formation of a well close to the nanofibre. This observation may pertain even when interatomic and fibre axes are not parallel.
We finally showed that the presence of the fibre causes new anisotropic features in the interaction between two P Rydberg rubidium atoms. In particular, the rotation symmetry around the interatomic axis is broken, and the dependence on the angle between interatomic and quantisation axes is reshaped by the presence of the fibre.

We believe the work presented in this article is merely a glimpse into the richness of Rydberg-atom interactions near  an optical nanofibre. It calls for a thorough and systematic investigation of the great wealth of possible configurations, including, for instance, the interaction between atoms in different states, or with arbitrary interatomic and quantisation axes. Besides its fundamental interest, such a study potentially holds applicative promises for quantum technologies. For instance the identification of interacting versus non-interacting -- and therefore blockading versus non-blockading -- configurations may pave the way to quantum devices, such as Bragg mirrors or gates, with highly interesting functionalized properties.

\end{document}